\documentclass[aps,prl,showpacs,superscriptaddress,twocolumn,floatfix]{revtex4-2}
\usepackage{dcolumn}   % needed for some tables
\usepackage{bm}        % for math
\usepackage{amsmath}
\usepackage[Euler]{upgreek}
\usepackage{float}
\usepackage{multirow}
\usepackage{booktabs}
\usepackage{textcomp}
\usepackage[Euler]{upgreek}
\usepackage{lipsum}
\usepackage{rotating}
\usepackage{tikz}
\usepackage[normalem]{ulem}

\usepackage[bookmarks=false]{hyperref}
\begin{document}
	\title{Heterogeneous interactions and polymer entropy decide organization and dynamics of chromatin domains}
	
	\author{Kiran Kumari}
	\affiliation{IITB-Monash Research Academy, Indian Institute of Technology Bombay, Mumbai, Maharashtra -  400076, India}
	\affiliation{Department of Biosciences and Bioengineering, Indian Institute of Technology Bombay, Mumbai 400076, India}
	\affiliation{Department of Chemical Engineering, Monash University, Melbourne, VIC 3800, Australia}
	%\email{}
	\author{J. Ravi Prakash}
	\affiliation{Department of Chemical Engineering, Monash University, Melbourne, VIC 3800, Australia}
	%\email{}
	\author{Ranjith Padinhateeri}
	\affiliation{Department of Biosciences and Bioengineering, Indian Institute of Technology Bombay, Mumbai 400076, India}

\begin{abstract}
Chromatin is known to be organized into multiple domains of varying sizes and compaction. While these domains are often imagined as static structures, they are highly dynamic and show cell-to-cell variability. Since processes such as gene regulation and DNA replication occur in the context of these domains, it is important to understand their organization, fluctuation and dynamics. To simulate chromatin domains, one requires knowledge of interaction strengths among chromatin segments. Here, we derive interaction strength parameters from experimentally known contact maps and use them to predict chromatin organization and dynamics. Taking  two domains on the human chromosome as examples, we investigate its 3D organization, size/shape fluctuations, and dynamics of different segments within a domain, accounting for hydrodynamic effects. Considering different cell types, we quantify changes in interaction strengths and chromatin shape fluctuations in different epigenetic states.
Perturbing the interaction strengths systematically, we further investigate how epigenetic-like changes can alter the spatio-temporal nature of the domains. 
Our results show that heterogeneous weak interactions are crucial in determining the organization of the domains. Computing effective stiffness and relaxation times, we investigate how perturbations in interactions affect the solid-like and liquid-like nature of chromatin domains. Quantifying dynamics of chromatin segments within a domain, we show how the competition between polymer entropy and interaction energy influence the timescales of loop formation and maintenance of stable loops. 
\end{abstract}
\maketitle

\section{Introduction}
Chromatin is a long polymer made of DNA and proteins; its four-dimensional (spatial and temporal) organization is crucial in regulating cellular processes like transcription, replication and DNA repair~\citep{bickmore2013spatial,pope2014topologically,wei2018three,sanyal2012long}.
How the chromatin self-organises spatially and temporally is a question of active research. 
Recent developments in chromosome conformation capture (Hi-C)~\cite{lieberman2009comprehensive,Dixon2012,Nora2012,nagano2017cell,sanyal2012long,bau2011three,shaban2020hi} and microscopy experiments~\cite{nozaki2017dynamic,bintu2018super,szabo2020regulation} help us investigate contacts among different segments and overall 3D organization of the chromatin polymer. Such experiments so far suggest that intra-chromatin interactions lead to the formation of different types of domains within a single chromosome~\cite{Dixon2012,Nora2012,Rao20141665,rowley2018organizational,bintu2018super,szabo2020regulation,belaghzal2021liquid}.

It has been hypothesized that loop extrusion, phase separation or an interplay between both could be the mechanism for the formation of these domains~\cite{alipour2012self,rowley2018organizational,mir2019chromatin,nuebler2018chromatin}. In the phase separation picture, interactions mediated by the different combinations of histone modifications and chromatin binding proteins lead to the formation of compartments/domains~\cite{mir2019chromatin,nair2019phase,hnisz2017phase}.  
In the loop extrusion picture, chromatin regions are actively brought together with the help of proteins like cohesins/condensins and held together by CCCTC-binding factor (CTCF)~\cite{Rao20141665,rowley2018organizational,goloborodko2016chromosome,fudenberg2016formation}.

While an interplay between loop extrusion and phase separation would determine the formation of many of the domains, it has been found that not all the domains have CTCF-dependent loops~\cite{Rao20141665,kaushal2021ctcf}. Recently, it has been shown that in the absence of loop extruding factors, chromatin does still form domains and execute the necessary biological function~\cite{mir2019chromatin,bintu2018super,benabdallah2019decreased,kaushal2021ctcf,gibson2019organization} indicating that micro phase-separation might be an important mechanism.
Phase separation is also known to bring together certain enhancers and promoters, segregating them from other regions~\cite{nair2019phase,hnisz2017phase,shrinivas2019enhancer}.
In certain cases, as far as biological function is concerned, there is an ongoing debate whether actual contact is crucial or proximity --- closeness in 3D without being physically in contact --- would suffice \cite{mir2019chromatin,benabdallah2019decreased,zhang2012spatial,wei2018three}. To probe this further, it is important to go beyond contacts and examine the whole configurational space of chromatin polymer.

Widely used experimental methods so far provide us only static snapshots of chromatin contacts. Large Hi-C contact values are sometimes assumed to represent static contacts holding together different segments of chromatin.
However, note that experimentally inferred contact probability values for nearly all chromatin segments are very small ($<<1$)~\cite{Rao20141665}. This implies that the contacts will be often broken, and the chromatin polymer conformation can be highly dynamic. Even in a steady state, there are likely to be large fluctuations, cell to cell and temporal variabilities. While there have been several attempts to understand the static 3D configurations and their average properties, the dynamic nature of chromatin domains remains unclear. 

Complementing recent experimental efforts, there have been several computational studies investigating chromatin organization~\cite{goloborodko2016chromosome,jost2014modeling,fudenberg2016formation,parmar2020nucleosome,kumari2020computing,giorgetti2014predictive,bianco2018polymer,tamar2019pnas,DiPierro2016,rosa2008structure,di20214d,clarkson2019ctcf,conte2020polymer,qi2019predicting,shi2018interphase,macpherson2018bottom,bajpai2020irregular,brackey2020mechanistic,ghosh2018epigenome,tjong2016population,hua2018producing,shinkai2020phi,tiana2015montegrappa,tiana2016structural,tortora2020chromosome,falk2019heterochromatin,di2016hi,buckle2018polymer}. 
Many studies carried out ``forward" simulations where the chromatin structure is predicted starting with a set of interaction strengths~\cite{brackey2020mechanistic,ghosh2018epigenome,shi2018interphase,macpherson2018bottom}. However, we do not know the optimal values of interaction strengths \textit{a priori}. One requires an inverse method that can obtain the optimal interaction strengths such that experimentally known properties are recovered. Studies have been carried out to optimise the interaction strength. Some groups optimise spring like interaction parameters while some others optimised LJ-like interaction parameters  ~\cite{giorgetti2014predictive,tiana2015montegrappa,zhang2015topology,meluzzi2013recovering,zhang2016shape,tokuda2012dynamical,tokuda2017heterogeneous,roux2013statistical,crehuet2019bayesian,cesari2018using,reppert2016refining} (see discussion section for more details).
We have recently developed an Inverse Brownian Dynamics (IBD) method to extract potential energy parameters that are consistent with HiC/5C experiments~\cite{kumari2020computing}. 
The method explicitly computes intra-chromatin interactions for every segment pair using an SDK potential. 
Using the intra-chromatin interactions computed from this method, in this work, we go beyond the static picture and study chromatin fluctuations and dynamics perturbing epigenetic states. 
We use Brownian dynamics (BD) simulations with hydrodynamic interactions (HI) for this study. Going beyond contact frequencies, we compute the full distance distribution between all pairs of segments and examine the cooperative nature of folding. We then study the dynamics of the domain, compute relaxation times,  loop formation times and contact times. We investigate how the interaction strengths and polymer entropy influence these measurable quantities. Finally, we discuss the significance of these findings.

\section{\label{sec:model} Model and Methods}
We consider chromatin as a bead spring chain having optimal intra-chromatin interactions derived from 5C and Hi-C data using an Inverse Brownian Dynamics (IBD) algorithm~\cite{kumari2020computing}. The total energy of the chromatin bead spring chain, made of $N$ beads, is $U=U^{\rm S} +U^{\rm SDK}$ where $U^{\rm S} =\sum_{i} \frac{H}{2}(|\bm r_{ i}-\bm r_{i+1}| - r_0)^2  $ is the spring potential between the adjacent beads $i$ and $(i+1)$,
$\bm r_{i}$ is the position vector of bead $i$, $r_0$ is the natural length and $H$ is the stiffness of the spring~\cite{bird1987dynamics}. To mimic protein-mediated interactions between bead-pairs, we use the Soddemann-Duenweg-Kremer ($U^{\rm SDK}$) potential whose repulsive part is modelled by the Weeks-Chandler-Anderson potential 
$4\left[ \left( \frac{\sigma}{r_{ ij}} \right)^{12} - \left(\frac{\sigma}{r_{ij}} \right)^6 + \frac{1}{4} \right] - \epsilon_{ij} $ upto the range $2^{1/6}\sigma$. 
Here $\sigma$ is the parameter that determines the minima of the potential, $ r_{ij}=|\bm r_{i}-\bm r_{j}|$ is the distance and $\epsilon_{ij}$ is an independent parameter representing the attractive interaction strength between beads $i$ and $j$. 
The repulsive part is unaffected by the choice of potential parameter $\epsilon_{ij}$.
The attractive part is modelled by $ \frac{1}{2} \epsilon_{ij} \left[ \cos \,(\alpha r_{ij}^2+ \beta) - 1 \right]$ which, unlike the Lennard-Jones potential, smoothly reaches zero at the cut off radius $r_c = 1.82\sigma$~\cite{soddemann2001generic,santra2019universality}.  Parameters $\alpha$ and $\beta$ control $r_c$ (see Table~S1 in supplementary information (SI)). We simulate the chromatin polymer using Brownian dynamics simulations, where the time evolution of the bead positions are governed by an It\^{o} stochastic differential equation. This simulation accounts for hydrodynamics interaction which is computed by the regularized Rotne-Prager-Yamakawa tensor. (See sec.~S1)
For the simulation, all the length and time scales are non-dimensionalised with $l_H=\sqrt{k_{\rm B}T/H}$ and $\lambda_H=\zeta/4H$, respectively where $T$ is the absolute temperature, $k_{\rm B}$ is the Boltzmann constant, and $\zeta=6\pi\eta_s a$ is the Stokes friction coefficient of a spherical bead of radius $a$ with $\eta_s$ being the solvent viscosity. 
All non-dimensional quantities are indicated with the asterik (*) symbol such as the non-dimensional position $r_{i}^* = r_{i}/l_H$. 

For chromatin, we do not know the interaction energy strengths {\it \`a priori}. Given 5C/HiC data, we performed an inverse calculation and obtained the optimal interaction strengths ($\epsilon_{ij}$) between bead $i$ and bead $j$, that are consistent with the experimental contact probability map~\cite{kumari2020computing}. The inverse algorithm works as follows: we start with the initial guess values of interaction strengths, simulate the polymer following the conventional forward Brownian dynamics method and obtain the simulated contact probabilities in the steady-state (IBD flow chart is shown in Fig.~S1).
Based on a statistical method, the interaction strengths ($\epsilon_{ij}$) are revised for the next iteration, depending upon the difference between the simulated and the known experimental contact probabilities (see sec.~S1).
We perform several iterations of the loop (i.e., BD simulation, calculation of contact probability and revision of  $\epsilon_{ij}$) until the error between the simulated and experimental contact probabilities is less than a predetermined tolerance value. 
At the end of the IBD process, we obtain the optimised interaction strength values ($\epsilon_{ij}$) that are independent of the initial guess values (also see~\citet{kumari2020computing}).
For each iterative step, we have generated 1000 trajectories and collected 100 samples from each trajectory, making it an ensemble of $10^5$ configurations.
Some of the earlier works have used the maximum entropy approach to obtain an optimal energy function. In a sense, the procedure we have applied can also be considered as a maximal entropy approach as the Brownian dynamics simulations give us the maximum entropy distribution --- ``equilibrium'' distribution  or steady state distribution -- given a set of constraints.  

In this manuscript, we have studied two different loci in the human genome. We primarily focus on
 human $\alpha$-globin gene locus (Chr:5, ENm008 region); we have simulated it starting with the experimental data available from~\citet{bau2011three}, in two different cell types: in the GM12878 cell type where the $\alpha$-globin gene is in its repressed state (throughout the manuscript we will refer to this state as the ``OFF" state), and in the K562 cell type where the gene is in its active state (throughout the manuscript we will refer this state as the ``ON" state).
 In this work, we have coarse-grained 10kb chromatin as one bead and the size of the bead is estimated to be 36nm see SI for details). We simulated the polymer for $\approx 10^4 \lambda_H$, where $\lambda_H$ is estimated to be 0.1sec (see SI).  
 The optimised $\epsilon_{ij}$ values for these states are shown in Fig.~S2 (also see~\citet{kumari2020computing}). We have repeated this study on a different locus (human Chr.7, a 500~kb region) for two cell types: IMR90 and K562. The different cell types for the same region represent different epigenetic states of the domain. 

\section{\label{sec:results}Results and discussion}
\subsection{\label{sec:res_eps} Prediction of interaction energies between chromatin segments}
As a first step, we have analysed the optimised interaction strengths ($\epsilon_{ij}$), predicted using our IBD method, for $\alpha$-globin gene locus for two different cell types (OFF and ON epigenetic states). The frequency distribution of the optimised $\epsilon_{ij}$ values is shown in Fig.~\ref{fig:1}(a).
In the ON state, the smaller values of $\epsilon_{ij}$ are more frequent, while the OFF state has a wide distribution of $\epsilon_{ij}$, suggesting a more compact state, a common feature of a heterochromatin/repressed state~\cite{boettiger2016super}. 

We then investigated the relation between the optimised $\epsilon_{ij}$ values (from IBD) and the contact probability values ($p_{ij}$) from experiments. We find that, for any given bead pair, contact probability value alone does not determine $\epsilon_{ij}$. Studying the data in detail, we propose the following empirical relation that can predict the interaction strength of a given bead pair:
\begin{equation}
\epsilon_{ij} = \epsilon_{0} + \alpha \log{\left( \frac{p_{ij}}{p_{min}(i-j)} \right)},
\label{eq-epsilon}
\end{equation}
where $p_{min}(i-j)$ is the minimum non-zero value of the contact probability of all beads having genomic distance $|i-j|$, and the other two parameters $\alpha \approx 1$, $\epsilon_{0}\approx 0$ are numerical constants (see~S2). In Fig.~\ref{fig:1}(b) \& (c), we present the comparison between the $\epsilon_{ij}$ predicted from the above empirical relation and the optimised $\epsilon_{ij}$ from IBD for ON and OFF states respectively. One can see that the formula predicts the $\epsilon_{ij}$ values reasonably well. The $\epsilon_{ij}$ from IBD simulation has larger spread; however, both have a similar behavior and very close numerical values. See Fig.~S3 and sec.~S2 for more details. This formula would greatly help in generating excellent initial guess values of $\epsilon_{ij}$ and will lead to faster convergence of the IBD. 

Next, we computed the average 3D spatial distance ($r_{ij}^* = |{\bf r}_{i}^*-{\bf r}_{j}^*|$) between different pairs of chromatin segments as a function of the corresponding 1D genomic separation ($s_{ij}=\lvert {i}-{j}\rvert$) for $\alpha$-globin gene (OFF state), and have compared with a ``control'' simulation of a self-avoiding walk (SAW) polymer with no attractive interaction ($\epsilon_{ij} =0$). While the SAW results in $\nu=0.6$ as expected, the OFF state has a scaling exponent $\nu=0.38$ suggesting a near closed packing (see Fig.~S4 and sec.~S3). 
Depending on the organism and epigenetic state, one might observe different values for exponent $\nu$ as experimentally seen by Zhuang lab~\cite{boettiger2016super}. 
3D distance computations help us to fix parameters in the model. 
We used published FISH data for distance between a single pair of $\alpha$ globin  segments and deduced $l_H=36$nm, and using the estimates of diffusion time scales we deduced $\lambda_H = 0.1$ (see Fig.~S5 and sec.~S4). Throughout this paper, $l_H=36$nm and $\lambda_H=0.1$s are used to convert all non-dimensional lengths and times, respectively, into standard units. We will present quantities in both non-dimensional and dimensional units. The reasons for the choice of both these specific values are discussed in greater detail in sec.~S3 and S4.

\subsection{\label{sec:res_prob} Distance distributions and cooperative nature of chromatin folding}
Even though the average distance between two chromatin segments is often used to represent chromatin organization, this may not describe the accurate biological picture in a dynamic, heterogeneous context. We compute the distribution of the 3D distance between different segments, $p(r^*)$, as it captures the maximum information about variability and fluctuations of chromatin. 
As a control, we computed the $p(r^*)$ for a SAW and it agrees well with the known analytical expression of des Cloizeaux, $p(\bm r^*) = C (r^{*})^{\theta +2} e^{-(K r^*)^{1/(1-\nu)}}$~\cite{des1980short} . Here $\nu$ is the Flory exponent, $\theta$ is a geometrical exponent and the coefficients $C$ and $K$ are functions of $\theta$~\cite{des1980short}. 
Recent work has led to an accurate estimation for these constants which are discussed in sec.~S5.
Fig.~\ref{fig:1}(d) shows the validation with intermediate beads of a SAW (with $\nu=0.6, \theta=0.81, K= 1.17, C=2.05$; also see Fig.~S6). 

\begin{figure*}
	\begin{center}
	{\includegraphics*[width=0.8\linewidth]{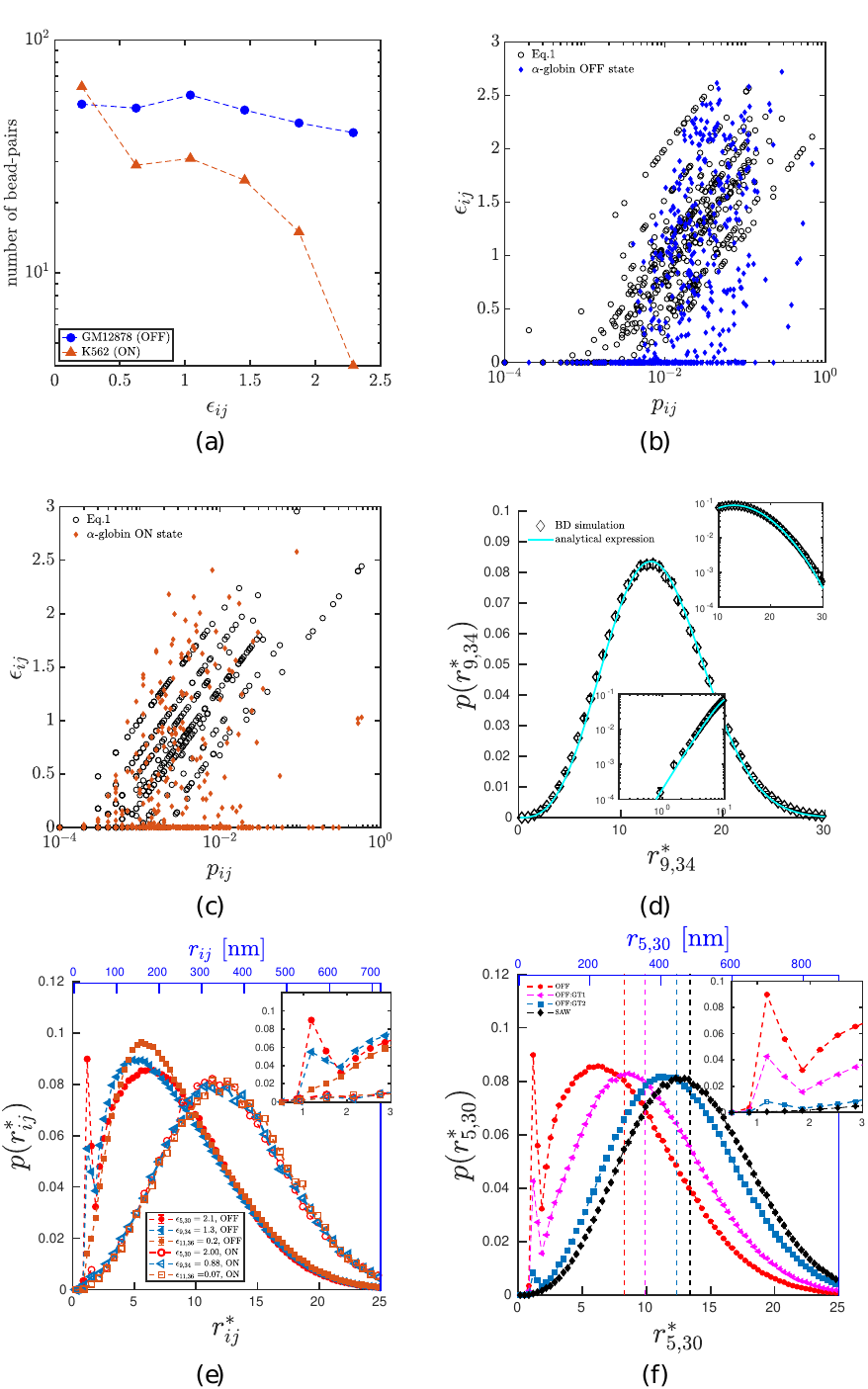}}
	\end{center}
\end{figure*}

\begin{figure}[h!]
		\caption{ 
			{\bf Distribution of interaction strengths and 3D distances}:
			(a) Distributions of the optimised interaction strengths ($\epsilon_{ij}$) for the ON and OFF states of the $\alpha$-globin gene locus. The ON state is populated with smaller values of $\epsilon_{ij}$ while the OFF state has nearly a flat distribution with similar occurrence of all values of $\epsilon_{ij}$.
			(b-c) Prediction of the relation between interaction strength and contact probability from the IBD simulation(filled diamonds) and from the eq.~\ref{eq-epsilon}(black open circles) for the  OFF state(b), and ON state(c) of $\alpha-$globin.
			(d) Distance probability distribution $p(r_{ij}^*)$ from simulations compared with the analytical des Cloizeaux expression~\cite{des1980short} for a pair of beads $9$ and $34$ of a SAW polymer of total length $N=50$. Insets show the zoomed in view of the fit for small (bottom inset) and large (top inset) values of $r^*$. 
			(e) $p(r_{ij}^*)$ for various bead pairs with different interaction strengths $\epsilon_{ij}$, but same $s_{ij}=25$ in the the OFF state (filled symbols) and the ON state (empty symbols). Inset: the small $r^*$ regions is zoomed in the inset to highlight the interaction-driven peak in the OFF state.  
			(f) Comparison of $p(r_{ij}^*)$ between different perturbed epigenetic-like states. OFF: state with all interactions in GM12878 (red), OFF GT1: when weak interactions are ignored; i.e., only with $\epsilon_{ij} > 1k_{\rm B}T$ interactions (pink), OFF GT2: when only very high interactions ($\epsilon_{ij} > 2k_{\rm B}T$) are accounted for (blue), SAW: control simulation with no attractive interaction (black). Vertical dashed lines represent  $\langle r_{ij}^* \rangle$ corresponding to each distribution. \label{fig:1} }
\end{figure}

\begin{figure*}
	\begin{center}
		\begin{tabular}{c}
			{\includegraphics*[width=0.7\linewidth]{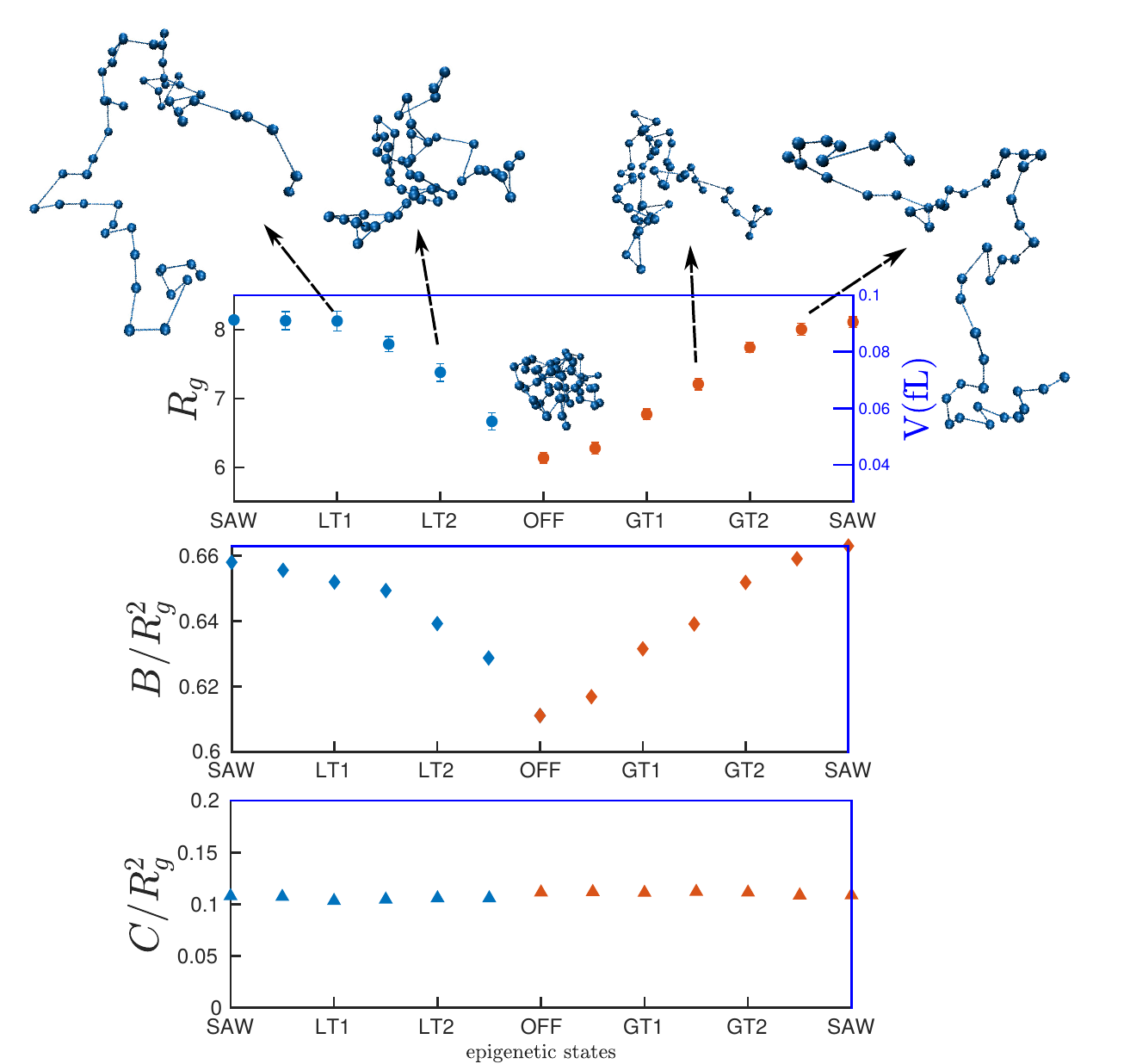}}\\
		\end{tabular}
		\caption{ {\bf Interaction strengths determining shape properties of chromatin domain:} Upper panel: Size of the chromatin domain ($R_g$) as we perturb interaction strengths. $x$-axis represents different interaction states with
			extreme ends representing the control (SAW) polymer and the OFF state in the middle. LT1 (LT$x$)  indicates that all interactions below 1$k_BT$ ($x k_BT$) are present in the polymer. Similarly GT1 (GT$x$) indicates that all interactions above 1$k_BT$ ($x k_BT$) are present. The right $y$-axis indicates volume of the chromatin domain in femtolitre. Snapshots from simulations at various epigenetic states are shown around the perimeter of the graph. Bottom two panels  represent the normalized asphericity ($B/R_g^2$) and acylindricity ($C/R_g^2$), respectively. See sec.~S7 for the definition of $B$ and $C$. The $x$-axis is the same in all panels. 
			 }
			 \label{fig:rg2}
	\end{center}
\end{figure*}

We then studied the $p(r^* )$ for the $\alpha$-globin gene locus in both GM12878 and K562 cell types. Examining various segments 250kb apart along the chain backbone ($25$ beads) for the OFF state, we find that all the distributions have a broad peak near their respective average distances (Fig.~\ref{fig:1}(e)), with an overall asymmetric tilt towards smaller $r^*$ values, when compared with SAW (See some sample snapshots in Fig.\ref{fig:rg2}). 
For bead pairs having high $\epsilon$ values, a sharp peak emerges near $r^* \approx r_c$ — we call this an “attraction-driven peak" as it is within the attractive range of the potential. The height of the peak is correlated with the strength of attraction ($\epsilon$). 
In an ideal homogeneous polymer case (SAW), we observe a single peak with no signature of attraction-driven peak. On the other hand, chromatin configuration is highly heterogeneous and the bimodal distribution of distances indicate the presence of more than a single population of structures. This is also analogous to the FISH and Hi-C paradox presented in \cite{shi2019conformational} where they observed multiple peak in 3D distance distribution as the result of heterogeneity. This difference is also reflected in the cumulative distribution function as shown in Fig.~S7. The heterogeneity information gets buried in ensemble averaged quantities such as average distance.

Similarly, we computed $p(r^*)$ for the ON state as well. Interestingly, we do not see the attraction-driven peak near $r^* \approx r_c$. The shape of the curve is similar to the SAW showing more symmetric distribution and independent of the identity of the bead-pairs -- it depends mainly on the genomic separation. 
Together, these results imply that average distances between bead pairs may not represent the complete picture of chromatin organisation; understanding the whole distribution is necessary. 

Given that we have the optimal interaction strengths that satisfy the experimentally known contact probability constraints~\cite{kumari2020computing}, we can answer the following important question: Are the measurable properties of a given bead-pair (e.g. $r_{5,30}$) solely determined by the interaction between those two particular beads ($\epsilon_{5,30}$) or are they influenced by the interactions among other bead-pairs as well?
To answer this, we adopted the following strategy: we systematically switched off the attractive interaction among certain bead pairs and computed probability distributions and other polymer properties. We simulated polymers for the following four cases: (i)~all interactions are considered -- GM12878 (OFF), (ii) only those interactions above $1 k_BT$ are considered -- we call it OFF:GT1 -- all weak interactions ($<1 k_BT$) are switched off here, (iii) only strong interactions above $2 k_BT$ are considered -- OFF:GT2 -- all weak and medium interactions ($<2 k_BT$) are switched off (iv) all attractive interactions are switched off -- the SAW polymer. These four cases can be imagined as four different epigenetic-like states -- states having different interaction strengths due to underlying epigenetic variations (see sec.~S6 and Fig.~S8).  
Fig.~\ref{fig:1}(f) shows $p(r_{5,30})$ for all the four cases. When we switch off the weak interactions below $1 k_BT$ (OFF:GT1), compared to the OFF state, the height of the interaction-driven peak of the distribution decreases and overall, the polymer swells resulting in the shift of the second peak (compare pink and red curves in Fig.~\ref{fig:1}(f)). This implies that weak interactions having strengths comparable to thermal fluctuations can also influence the contact probability and polymer configurations. If we keep only the highly prominent interactions and neglect all interactions below $2k_BT$ (OFF:GT2), the interaction-driven peak further diminishes, and the distribution function approaches the SAW distribution (compare blue with other curves in Fig.~\ref{fig:1}(f)). Note that the interaction between beads 5 and 30 is present ($\epsilon_{5,30} = 2.09$) in all the cases except in the SAW case.
These results suggest that the measurable properties for a given bead-pair (e.g. $r_{5,30}$) depends not only on the attraction strength of that particular bead pair but also on the interactions of the whole polymer chain. This result implies that all bead pairs collectively/cooperatively contribute in determining the relative position for a particular bead pair. 

\subsection{Relevance of weak and strong interactions in altering the volume and shape of the chromatin domains}

The above picture suggests that the chromatin folding is influenced by the collective behaviour of all beads having different interaction strengths. 
To examine the nature of collective behaviour, we probed a property of the whole polymer, namely the radius of gyration ($R_g$) defined in sec.~S7. 
To understand how folding is affected by different epigenetic states, we did the following. We started with a polymer having no interactions (SAW), added weak interactions (small $\epsilon$) that exist between beads in the OFF state as the first step, equilibrated, computed $R_g$ and sequentially added stronger interactions between beads step by step ($\epsilon<0.5, \epsilon<1.0, ..., \epsilon<2$ and so on, denoted as LT1, LT2 etc.), until the OFF state (GM12878) is reached. Each step was equilibrated and $R_g$ was computed (see Fig:\ref{fig:rg2}, top panel). 
From $R_g$, we have also computed the volume $V= (4/3) \pi R_g^3$ of the chromatin domain as shown in the right side $y$-axis. 
As seen from the figure, adding very weak interactions does not change $R_g$ much. However, adding intermediate interactions significantly reduces the $R_g$, and it saturates as the interactions get stronger, resulting in a sigmoidal-like curve showing signatures of cooperative/collective behaviour.  
Since we have equilibrated the polymer for each set of $\epsilon$ values, the LHS of the curve can be interpreted in two ways: folding the polymer by adding stronger and stronger interaction starting with a completely unfolded state or equivalently unfolding the polymer by removing the stronger interaction starting with a completely folded OFF state. 
One can also ask how the polymer would fold if one adds strong interactions (larger $\epsilon$) as the first step, starting with SAW, and then add weaker interaction sequentially step by step (denoted as GT1, GT2 etc.). This is shown in the RHS of Fig.~\ref{fig:rg2} (orange symbols). 
The whole curve suggests that having prominent interactions alone or weaker interactions alone may not take the system closer to its full equilibrium state.  
We also show typical snapshots of 3D chromatin configurations corresponding to different epigenetic states. As expected, the OFF state is compact, and the volume of the domain increases as we go towards the SAW state. The approximate two fold-change in volume between the two extreme states seen here are roughly the same order as the density change observed experimentally~\cite{imai2017density}. The predicted values of the $R_g$ and volume are also of the same order of magnitude as reported by~\citet{boettiger2016super}.

To quantify how the shape of the chromatin domain changes with epigenetic states, we computed the asphericity (B)  and the acylindricity (C) parameters (see sec.~S7 for definition). Asphericity quantifies the extent of deviation from a spherical shape. If a polymer is coiled with the average shape of a sphere, $B=0$. Here a positive B value suggests that even in the OFF state, the chromatin domain is not a perfect sphere. 
As we go from OFF to SAW, the asphericity increases by $\approx$65\% as shown in sec.~S7. However, the asphericity scaled with the polymer size ($B/R_g^2$) changes by $\approx$10\%. 
Similar to $R_g$, we have shown the GT (RHS, orange symbols) and LT (LHS, blue symbols) cases for the asphericity too. Even though both sides are monotonically increasing, note that LT cases are not equivalent to the GT cases. 
We also compute the acylindricity parameter that quantifies the extent of the deviation from a perfect cylinder. Here too, $C>0$ values suggest that the chromatin domain is not a perfect cylinder (see the lower panel of Fig.~\ref{fig:rg2} and sec.~S7).
Even though the acylindricity is monotonically increasing (shown in Fig.~S9) from the OFF state to a SAW, it is increasing in proportion to the size of the polymer. Hence the scaled acylindricity ($C/R_g^2$) is nearly a constant, as shown in the lower panel of Fig.~\ref{fig:rg2}. 

\subsection{Weak interactions are crucial in recovering 3D organization of chromatin}
\begin{figure*}
	\begin{center}
			{\includegraphics*[width=1.0\linewidth]{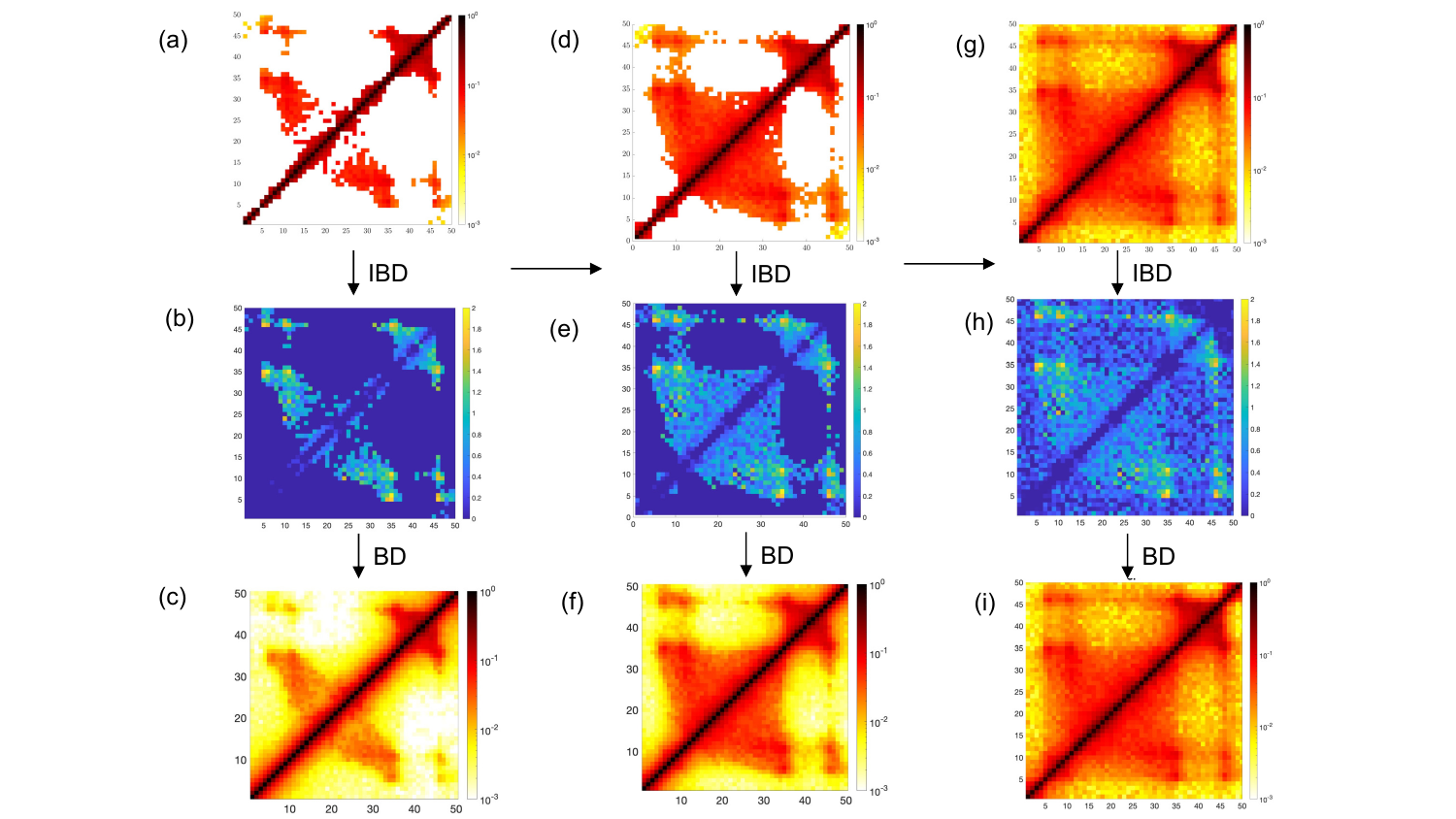}}
		\caption{{\bf Step-wise IBD showing the importance of weak interactions:} Stepwise IBD process for the Chr7 region in the IMR90 cell line. The IBD optimisation is presented in 3 steps.  (I) First column: input is only the prominent probability values (higher than one standard deviation from the mean. i.e., $p_{ij}>\bar{p}_{|i-j|}+\sigma_{|i-j|}$) show in (a) based on the peak detection algorithm (see text and S8). (b) The corresponding optimised $\epsilon_{ij}$ values, and (c) the recovered contact probabilities. (II) Second column: after the first optimisation step, all peaks above the average probability values (shown in (d)) are fed as input (i.e., $p_{ij}>\bar{p}_{|i-j|}$). The corresponding optimised interaction strengths in (e) are improved with some weak interactions appearing in this 2nd step. However note that the recovered contact probability simulated with prominent interactions in (f) is not comparable to  the full contact probability in (g).  (III) Third column: the complete Hi-C matrix in (g) is fed as input. At the end of this third step  the whole contact probability matrix was recovered in (i), and the corresponding optimal interaction strengths are predicted in (h).  \label{fig:ibd2} }
	\end{center}
\end{figure*}

To demonstrate our method and results on a different chromatin domain, we studied another 500~kbp domain (137,250~kbp-137,750~kbp) on human Chr7. Starting from Hi-C data (10kb resolution) from~\citet{Rao20141665}, we performed IBD and computed optimal interaction strengths (see details in sec.~S8) for cell types describing different epigenetic states (cell lines IMR90 and K562) for the same location. As the Hi-C data is denser compared to the 5C, we converged for the Hi-C contacts in a step-wise manner. In the first step (shown in Fig.~\ref{fig:ibd2} for IMR90), we optimized only for the stronger contacts (prominent peaks) in the Hi-C data. The prominent (strong) contacts are defined as follows: for each value of the $|i-j|$ line (line parallel to the diagonal representing all equidistant bead pairs), the mean and standard deviation for that off-diagonal line is computed as $\bar{p}_{|i-j|}$ and $\sigma_{|i-j|}$, respectively. If the contact probability for a matrix element in that $|i-j|$ off-diagonal line is greater than this mean + standard deviation ($p_{ij}>\bar{p}_{|i-j|}+\sigma_{|i-j|}$), it is considered as a prominent probability value. In the next step, we added weaker contacts such that all contacts greater than the mean ($p_{ij}>\bar{p}_{|i-j|}$) are present and achieved the IBD convergence. Finally, we included the rest of the contacts (all contacts) and obtained the optimal interaction strengths that can reproduce the complete experimental HiC map (Fig.~\ref{fig:ibd2}). It is interesting to note that in the first and second steps when weak contacts (contacts smaller than the mean) are not present, the contact probability from our simulations did not recover the complete contact map. This step-wise optimization process illustrates that the weak contacts are equally important, and the prominent contacts alone cannot reproduce the complete Hi-C map. This is consistent with what we found for the $\alpha$-globin domain in the earlier section. A similar analysis for K562 is shown in Fig.~S10, and a comparison between $\epsilon_{ij}$ for both the cell lines are shown in Fig.~S11. 

\begin{figure*}
\begin{center}
			{\includegraphics*[width=0.9\linewidth]{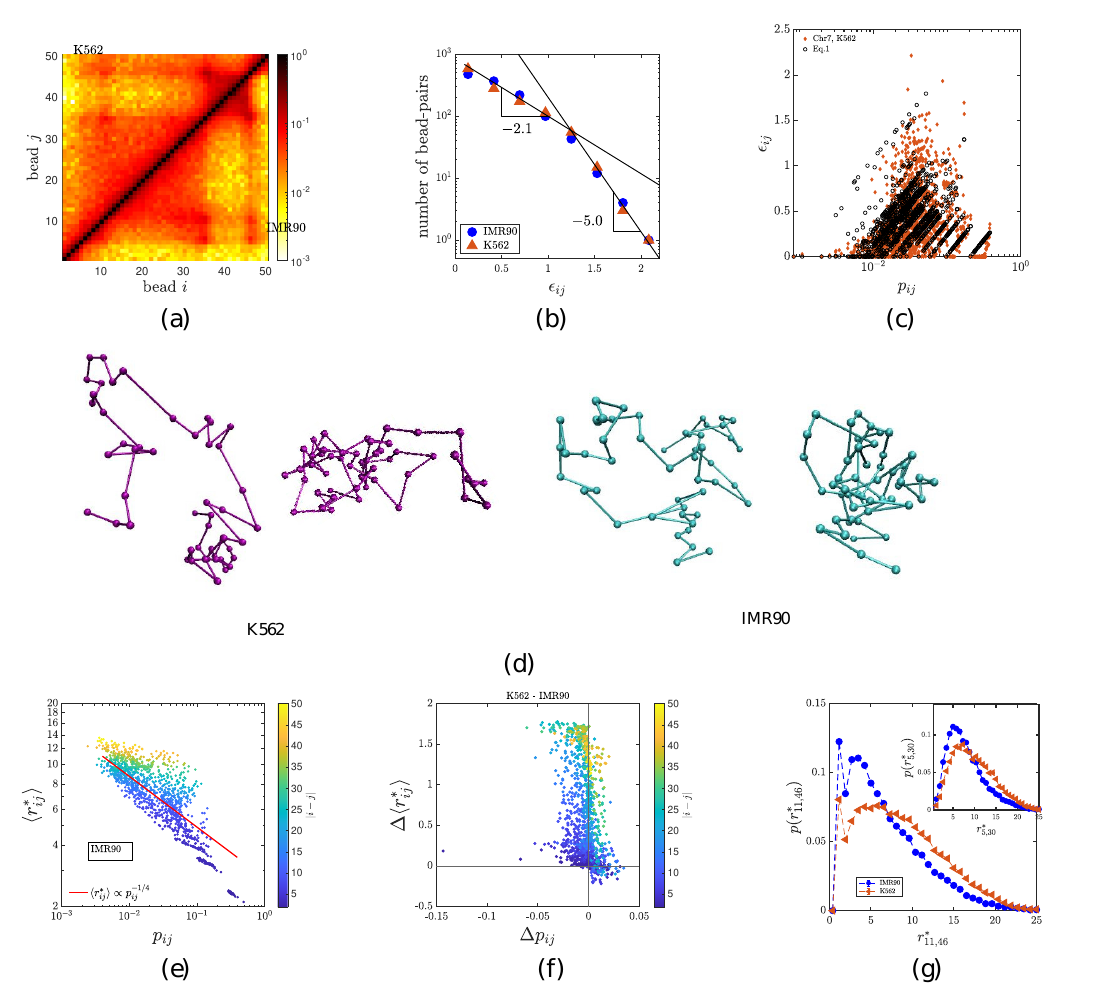}}
\end{center}
		\caption{{\bf Chromatin configuration and fluctuations for different epigenetic states}: 
		(a) The recovered contact probabilities ($p_{ij}$) for two epigenetic states/different cell types, K562 (upper triangle) and IMR90 (lower triangle) for a 500 kbp domain (137,250 kbp-137,750 kbp) on human Chr7. Contact maps for both the cell lines are roughly similar. However, the difference can be seen around certain bead pair like 10-45. 
		(b) The distributions of the optimised interaction strengths ($\epsilon_{ij}$) for the corresponding epigenetic states (cell types). Both the distributions are peaked at low values of $\epsilon_{ij}$ and are sharply decaying. Two exponential lines are given as a guide to the eye.
		(c) Prediction of the relation between interaction strength and contact probability from the IBD simulation(filled diamonds) and eq.~\ref{eq-epsilon}(black open circles) for the Chr.7 region from K562 cell line.
		(d) Two randomly selected  3D configurations for each epigenetic state (K562 and IMR90). 
		(e) Our prediction of the mean 3D distance between every pair of beads ($r_{ij}$) as a function of their corresponding contact probability ($p_{ij}$) for one of the epigenetic states(IMR90). The color indicates genomic distance $|i-j|$ between the pair of beads (in units of bead size, see sidebar). Curve with a power-law relation $r_{ij} \propto p_{ij}^{(-1/4)}$ is plotted (red line)  as a guide to the eye.
		(f) The difference between K562 and IMR90 epigenetic states. $\Delta r_{ij}=r_{ij}$(K562)$-r_{ij}$(IMR90) is predominantly positive suggesting that K562 is slightly more extended (see text).
		(g) Probability distribution of 3D distance, $p(r_{ij}^*)$, for bead pairs $11-36$ (main figure) and $5-30$ (inset). The former pair has high contact probability and shows a double peak. The latter (inset) is for a randomly selected pair. Both the main figure and inset suggest that K562 is slightly more extended than IMR90.   ~\label{fig:chr7}}
\end{figure*}
\subsection{Distribution of interaction strengths, 3D organization and fluctuations of chromatin domain for different epigenetic states}
Now that we have used the IBD method and obtained optimized $\epsilon_{ij}$ for the new domain, we can analyze the statistical properties of $\epsilon_{ij}$, use $\epsilon_{ij}$ values in a forward simulation, and obtain 3D organization of the domain for both the epigenetic states (IMR90 and K562).
First, we present contact probability recovered from our simulation (Fig.~\ref{fig:chr7}(a)) for both the epigenetic states (upper triangle is K562 and the lower triangle is IMR90). Since the IBD algorithm iteratively reduces the difference between the input HiC contact probability (experimentally obtained) and the recovered contact probability, this map from the simulation is identical to the contact map from Hi-C up to a very low tolerance (see Method). Note that contact maps for both the cell types are roughly similar. However, there are some important differences for some of the contact pairs (e.g., 10-45). We aim to examine whether these small differences in contact probability would lead to detectable differences in 3D distance and other quantities.
In Fig.~\ref{fig:chr7}(b), we present the distribution of optimized interaction strength for both the cell lines (IMR90 and K562). All the distributions are peaked at smaller values and decrease gradually, reminiscent of exponential decay. The interaction strength per bead-pair is in the range 0-2 $k_{\rm B}$T. Here too we used eq.\ref{eq-epsilon} and predicted the $\epsilon_{ij}$ values. As seen from Fig.~\ref{fig:chr7}(c), the prediction provides a reasonable comparison with what is obtained from the IBD simulation, indicating that the eq.\ref{eq-epsilon} works well for different cell types and different experimental datasets (more details in sec.~S2).

We now present representative 3D configurations for both the cell lines in Fig.~\ref{fig:chr7}(d). To quantify the shape and size of this region, we computed $R_g$, asphericity($B$) and acylindicity($C$). As represented in Table I, $R_g$ shows that K562 is more extended compared to IMR90. In other words, the epigenetic changes, appearing as small differences in contact maps and $\epsilon_{ij}$, are affecting the 3D configuration of chromatin in these cell types differently. The parameters $B$ and $C$ suggest that IMR90 is more spherical and K562 is more cylindrical compared to each other. 
\begin{table*}[]
\caption{Various shape properties based on the eigenvalues of the gyration tensor G are described here for cell lines IMR90 and K562 for the domain we studied in chromosome 7. See sec.~S7 for definition of these symbols.}
\label{tab:my-table}
\begin{tabular}{|c|c|c|}
\hline
Shape property & cell line IMR90   & cell line K562    \\ \hline
$R_g^2$        & 30.25   & 40.20   \\ \hline
$B$            & 17.50 & 25.67  \\ \hline
$B/R_g^2$      & 0.54   & 0.64   \\ \hline
$C$            & 3.52   & 4.20  \\ \hline
$C/R_g^2$      & 0.116   & 0.104   \\ \hline
prolateness    & -0.970 & -0.974 \\ \hline
anisotropy     & 0.478 & 0.538  \\ \hline
\end{tabular}
\end{table*}

Our simulations can predict the precise relation between contact probability of any segment pair and its corresponding average 3D distance. Fig.~\ref{fig:chr7}(e) shows this relation for the domain we studied from the cell line IMR90 (see Fig.~S12 for K562). Interestingly, it shows a broad distribution of 3D distance for a given contact probability, similar to what is observed in $\alpha$-globin gene locus \cite{kumari2020computing}. Even though a single power law curve cannot describe the whole data, the fit to the broad data gives the relation $r_{ij} \sim p_{ij}^{-1/4}$. While the average behaviour of the curve is consistent with the earlier work~\cite{shi2021hi,wang2016spatial}, our simulations uniquely predicts the variability in the 3D distances. In Fig.~\ref{fig:chr7}(f), we show how the contact probabilities ($\Delta p_{ij}=p_{ij}$(K562)$-p_{ij}$(IMR90)) and corresponding 3D distances ($\Delta r_{ij}=r_{ij}$(K562)$-r_{ij}$(IMR90)) differ between both the cell types. The $p_{ij}$ differences are small. However, a small change in $\Delta p_{ij}$ ($\approx 0.05$) can change the 3D distance by $\approx$ 2units ($\approx60$nm) for certain pair of beads (beads having high $|i-j|$, green/yellow color). 
Intuitively, when $\Delta p_{ij}$ is positive, the $\Delta r_{ij}$ should be negative and vice versa. This is clearly seen in the second and fourth quadrants. 
The first quadrant, where both $\Delta r_{ij}$ and $\Delta p_{ij}$ are positive points to those bead pairs that are compacted/extended due to possible collective effects arising from other beads nearby.

Since a wide range of 3D distances are possible for any given contact probability, we computed the whole 3D distance distribution $p(r^*)$ between different segments for both the cell type (see Fig.~\ref{fig:chr7}(g)). For certain segment pairs, here too, we observe double-peaked distributions, showing attraction-driven peak as in the case of the $\alpha$ globin OFF state. For certain other segment pairs, the distribution has only a single peak, as shown in the inset of Fig.~\ref{fig:chr7}(g). However, in all the cases, the distributions are skewed towards the lower values of $r_{ij}$ suggesting more compaction compared to SAW. It is also evident that the K562 is more extended than IMR90 once again, showing that small changes in contact probability can have interesting differences in 3D distances.

\subsection{Prediction of probability of triple contacts} 
\begin{figure*}[ht] 
	\begin{center}
			{\includegraphics*[width=1.0\linewidth]{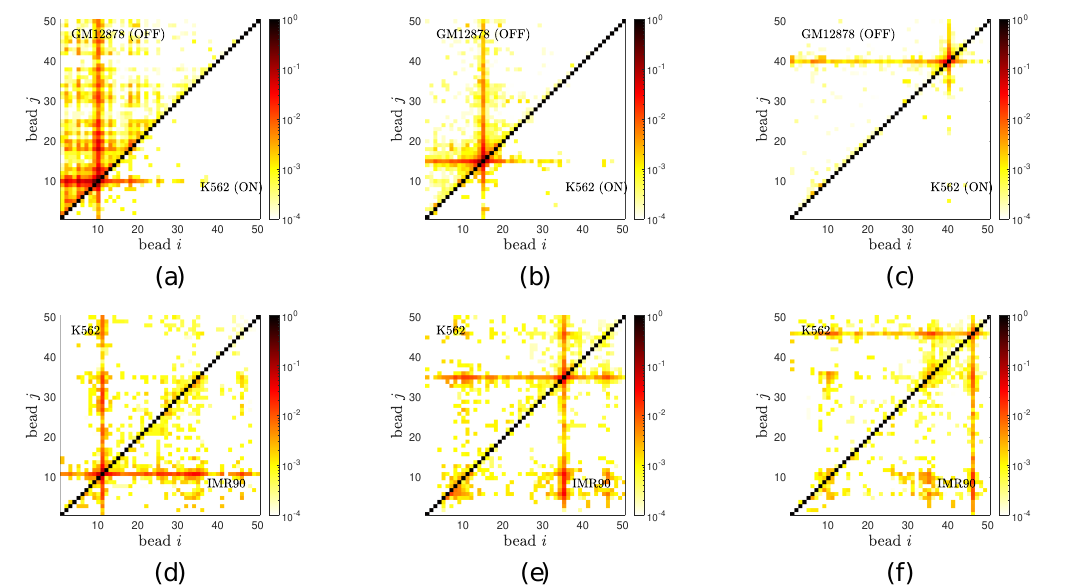}}
		\caption{{\bf Triple contact probabilities}: Probability of contact between 3 chromatin segments simultaneously, computed from simulations. (a) Triple contact probabilities of bead 10 (reference point) with any other two beads for the $\alpha$-globin gene locus in the OFF state (upper triangle) and ON state (lower triangle). (b) \& (c) are similar to (a) with reference points as beads 15 and bead 40, respectively. (d) Triple contact probabilities of bead 11 (reference point) with any other two beads on the Chr.7 region for K562 (upper triangle) and IMR90 (lower triangle).  (e) \& (f) are similar to (d) with reference points as beads 35 and bead 46, respectively.~\label{fig:triplec}}
	\end{center} 
\end{figure*}

Experimental contact probability data such as 5C and Hi-C can only provide pair-wise contacts. However, our simulations can go beyond pair-wise interaction and can provide multi-body contacts. Here in this section, we investigate three-body contacts --- simultaneous contacts between three beads in our simulations --- for various segments in different cell types. That is, we compute the probability of three beads (at positions $r_i,  r_j$ and $r_k$) coming together within a distance less than the cutoff radius (i.e. $r_{ij}<r_c$, $r_{ik}<r_c$,  and $r_{jk}<r_c$). 
 First, we investigate the triple contacts within the $\alpha$-globin domain. 
Figs.~\ref{fig:triplec}(a), (b) and (c) show the triple body contact with bead number 10, 15 and bead 40 as reference points, respectively. That is, the probability that bead 10 is in contact with any other two beads ($i$ and $j$ represented in $x$ and $y$ axis, respectively) is shown in Fig.~\ref{fig:triplec}(a) for both the ON (K562) and OFF (GM12878) states in lower and upper triangles of the matrices, respectively. Since, bead 10 is the reference point ($k = 10$), the vertical lines at beads 10 ($i = 10$) indicates the triple contact among $k = 10$ (the reference point), $i=10$, and $j$. Since the reference and $i$ are the same, this reduces to a pair-wise contact. Similarly, the horizontal line at bead 10 indicates the contact among $k = 10$ (the reference point), $i$, and $j=10$. 
All other matrix elements show the triple contact that bead 10 makes with different pairs of beads ($i$ and $j$). 
Interestingly, the OFF (GM12878) state has many more triple contacts as compared to the ON (K562) state. This is consistent with our earlier observation that the OFF state is highly compact while the ON state is close to SAW polymer. 
We repeated this analysis for different beads in the domain of our interest in Chr7. Fig.~\ref{fig:triplec}(d)-(f) show the triple contact with beads 11, 35 and 46 as reference points, respectively. Here too, we represent the K562 and IMR90 in upper and lower triangles of the matrices, respectively.  
In all three cases, IMR90 shows more triple contacts compared to the K562 cell line, specifically in the region bead 30 to bead 40. 

The biological significance of triple contacts is just emerging. There are new findings about ``shadow enhancers" where two enhancers interact with a single promoter~\cite{kvon2021enhancer}. One of the models for such interaction is the ``simultaneous looping model'' where two enhancers are simultaneously in contact with a gene.  There are also findings of one enhancer simultaneously regulating two different promoters. Computation of triple contact probabilities will help us understand the frequency of such events. A recent work has also suggested that high triple contact points indicate the boundaries of TADs~\cite{shi2022method}. 

\subsection{Estimation of stiffness and drag properties from domain relaxation times and  fluctuations}
\begin{figure*}[ht] 
	\begin{center}
			{\includegraphics*[width=0.9\linewidth]{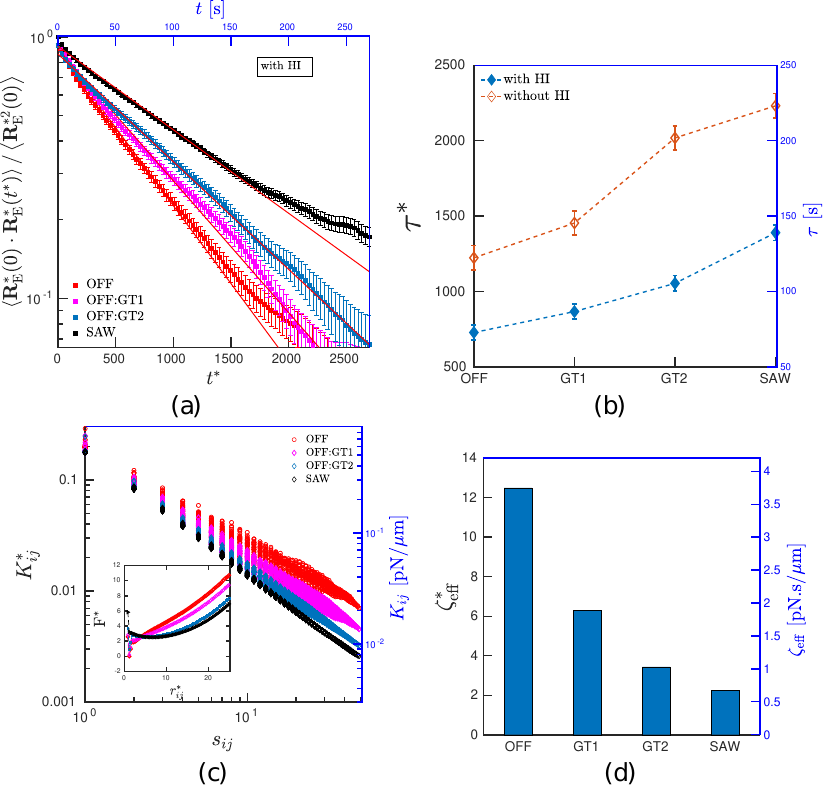}}
		\caption{ {\bf Stiffness and relaxation time characterising domain properties:} 
			(a) Exponential decay of end-to-end auto-correlation function with time for four epigenetic states computed with HI. 
			(b) Relaxation times ($\tau$) with and without HI reveal that HI helps in relaxing the polymer faster. 
			(c) Effective stiffness between all the bead-pairs; OFF state is more stiff compared to less interacting states and SAW. Inset: Free energy as a function of bead pair distance $r_{5,30}$. 
			(d) Effective viscous drag felt by different chromatin states. OFF state chromatin (with stronger interactions) is more stiff and has a higher viscous drag.
			\label{fig:3} }
	\end{center}
\end{figure*}
Whether chromatin is liquid-like, solid-like or gel-like has been a matter of intense discussion in the recent literature~\cite{maeshima2020fluid,nozaki2017dynamic,gibson2019organization,strickfaden2020condensed}.
In the phase separation picture, chromatin segments are thought to be “liquid-like", dynamically exploring various configurations. Given that our model can study the stochastic nature of formation and breakage of bonds, and polymer dynamics, consistent with what is observed in Hi-C experiments, below we compute relaxation times and fluctuations of the chromatin domain and estimate effective elastic and drag properties. To demonstrate our results, we use $\alpha$-globin domain as an example.

First, we computed the end-to-end vector autocorrelation function $\left<\mathbf{R}_{\mathrm{E}}^*(0)\cdot\mathbf{R}_{\mathrm{E}}^*(t^*)\right>/\left<\mathbf{R}^{*2}_{\mathrm{E}}(0)\right>$ where $\mathbf{R}_{\rm E}^* = (\mathbf{r}_1^*-\mathbf{r}_{50}^*)$ and extracted the longest relaxation time $\tau^*$ with and without hydrodynamic interactions (HI). 
The autocorrelation decay computed with HI is shown in Fig.~\ref{fig:3}(a) and no-HI case is shown in Fig.~S13. 
Fig.~\ref{fig:3}(b) shows that the relaxation times for all the perturbed states are lower with HI, as observed previously for the protein folding simulations~\cite{pham2008brownian,pham2010collapse}.
We also computed the $R_g$ autocorrelation function and observed similar behaviour.
All results presented in this paper are computed with HI unless stated otherwise.
The chromatin in the OFF state has a lower relaxation time compared to SAW chromatin (See Sec.~S9 for more details). Experiments measuring mean square displacement (MSD) have reported that repressed chromatin regions have an unexpected higher diffusivity compared to the non-repressed chromatin~\cite{germier2017real,nozaki2017dynamic}; considering that diffusivity is inversely proportional to the relaxation time, there is some similarity between the experimentally reported results and our findings.
 To understand this behaviour, we investigate the elastic and drag properties of the chromatin domain. From the measurement of fluctuations of each bead-pair we can compute an effective stiffness defined as $K_{ij}= k_{\rm B}T/\langle |\bf{r}_{\it i} - \bf{r}_{\it j}|^2\rangle$ (Fig.~\ref{fig:3}(c)). As expected, the highly interacting OFF state is stiffer than the other epigenetic states, including SAW. This can also be understood from the free energy as a function of bead pair distance $F^*(r^*)=-\ln{(p(r^*)/4 \pi r^{*2})}$ (see inset). 
The above behaviour is consistent with $K_{ij}^* \sim \dfrac{\partial^2 F^*_{ij}}{\partial r_{ij}^2}$
and stiffness ($K_{ij}^*$) of different epigenetic states do show similar behaviour. Since timescales in such problems are inversely proportional to the stiffness, the observed lesser time is explained by the higher stiffness.
For the known stiffness and relaxation times, we can compute an effective drag coefficient defined as $\zeta_{\rm eff}^*=\tau^* \times K^*$. 
Taking the effective stiffness of the end beads ($K_{\rm 1,50}^*$), we find that the drag for the OFF state is higher than the other states suggesting that the existence of larger attractive interactions reduces its ability to reorganize. 
Both the stiffness and drag are greater for the OFF state than the SAW, but they combine to lead to a faster relaxation time for the OFF state. 
Our findings are consistent with the recent experimental report that highly interacting chromatin shows reduced mobility as measured by Fluorescence Recovery After Photobleaching (FRAP) technique, revealing the gel-like nature of chromatin~\cite{strickfaden2020condensed,gibson2019organization}.

\subsection{Interplay between interaction energy and polymer entropy influences the dynamics of chromatin domain}
While we have gained insights into steady-state fluctuations and distance distributions, how the interactions would affect chromatin dynamics can be further probed~\cite{amitai2018encounter,zhang2016first,ghosh2018epigenome}.
We know that contacts between chromatin segments are dynamic; proteins that form contacts bind and dissociate, resulting in stochastic formation and breakage of contacts.  
This opens up interesting questions: How long do two beads remain in contact (looped)? When loops break and beads diffuse away, how long does it take for the bead pairs to come back in contact? What are the factors (interaction strengths, polymer entropy etc.) dictating the phenomena of dynamic contacts?
\begin{figure*}
	\begin{center}
			{\includegraphics*[width=0.6\linewidth]{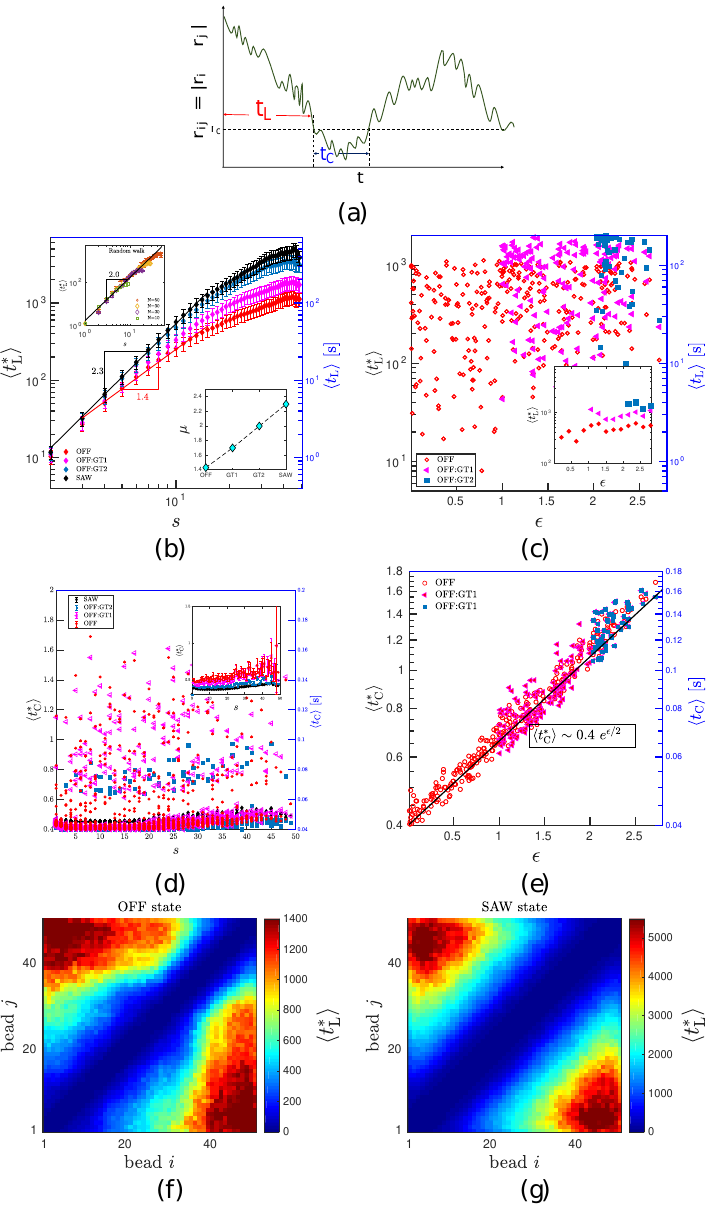}}
		 \end{center}
		\caption{{\bf Interaction energy and genomic seperation dictate the behavior of loop formation time $\langle t_{\rm L}^* \rangle $ and contact time $\langle t_{\rm C}^* \rangle $:}
			(a) schematic representation of the distance between two beads in a single trajectory showing $t_{\rm L}^*$ and $t_{\rm c}^*$. 
			(b) $\langle t_{\rm L}^* \rangle$ has a power law scaling with genomic length ($\langle t_{\rm L}^* \rangle \sim s^{\mu} $) with exponent varying from 1.4 (OFF state) to 2.3 (SAW) for different chromatin states. 
			The exponents are shown in lower inset.  The upper inset shows $\langle t_{\rm L}^* \rangle$ for various $N$ values for a random walk polymer. 
			(c) $\langle t_{\rm L}^* \rangle$ as a function of interaction strength with each point representing a bead pair. Note the huge spread in $\langle t_{\rm L}^* \rangle$.
			Inset: $\langle t_{\rm L}^* \rangle$ binned and averaged over all bead pairs having same $\epsilon$ showing minimal influence of $\epsilon$. 
			(d) $\langle t_{\rm C}^* \rangle$ as a function of $s$ with each point representing a bead pair. Here too, note the spread. Inset: $\langle t_{\rm C}^* \rangle$  binned and averaged over all bead pairs having the same $s$ showing minimal dependence on the  segment length.
			(e) $\langle t_{\rm C}^* \rangle$ increases exponentially with the interaction strength.  
			$\langle t_{\rm L}^* \rangle$  for all the bead pairs as a heatmap for (f) OFF and (g) SAW states. 
			\label{fig:4} }
\end{figure*}

To study the temporal nature of chromatin, similar to some of the earlier work~\cite{ghosh2018epigenome,toan2006depletion}, we define loop formation time ($t_{\rm L}^*$) and contact time ($t_{\rm C}^*$) for all bead-pairs. $t_{\rm L}^*$ is defined as the time taken for a pair of beads to meet ($r_{ij}^* < r_{\rm C}^*$) for the first time, starting from a random equilibrium configuration. $t_{\rm C}^*$ is defined as the duration that the bead-pairs remain looped/in contact.  A schematic representation of a typical time trajectory of 3D distance indicating $t_{\rm L}^*$ and $t_{\rm C}^*$ is shown in Fig.~\ref{fig:4}(a) and the actual data from our simulation, as an example, is shown in Fig.~S14.
Corresponding average quantities are defined by $\langle t_{\rm L}^* \rangle$ and $\langle t_{\rm C}^* \rangle$, respectively. 

Two possible factors that can influence these temporal quantities are interaction strengths ($\epsilon$) and polymer entropy. Since two beads having a larger segment length between them will have higher entropy, it is expected that the time to come into contact is longer.  In other words, the time of looping is expected to be dictated by polymer entropy. To validate this hypothesis, we looked at $\langle t_{\rm L}^* \rangle$ as a function of the genomic length with and without HI. 

As shown in Fig.~\ref{fig:4}(b) $\langle t_{\rm L}^* \rangle$ monotonically increases with $s$ showing a power law behavior $\langle t_{\rm L}^* \rangle \sim s^{\mu}$. As a control, we matched our $ \langle t_{\rm L}^* \rangle$ results with the previously known exponents $\mu \approx 2.3$ for SAW and $\mu = 2.0$ for a random polymer (see top inset)~\cite{toan2008kinetics,toan2006depletion}. By simulating various chain lengths ($N=10, 20, ...$), we can infer that the deviation from the power law for large $s$ is due to finite chain effects (top inset). We have also computed $\langle t_{\rm L}^* \rangle$ for all the other epigenetic states revealing $ 2.3 > \mu \geq 1.4$. The OFF state having all interactions shows the smallest exponent of $1.4$. As we remove interactions from the system, $\mu$ gradually approaches the SAW limit. The scaling appears independent of HI as shown in Fig.~S15. The change in power law may also be understood by looking at the free energy plotted in Fig.~\ref{fig:3}(c) inset. One can see that the free energy has a higher tilt in the OFF state compared to the other states, implying that the bead-pairs can move along the landscape quicker in the OFF state. The results for $\langle t_{\rm L}^* \rangle$ suggests that even in the absence of loop extrusion, the looping time is not too long (seconds to minutes). This also indicates that the micro phase-separation could be a viable mechanism for bringing together chromatin segments and possibly explains the experimentally observed fact that chromatin is functional even in the absence of loop extruding factors~\cite{benabdallah2019decreased,bintu2018super,kaushal2021ctcf}. We then examined how the interaction strength influences $\langle t_{\rm L}^* \rangle$, and found that there is a huge spread in the $\langle t_{\rm L}^* \rangle$ values, for a given $\epsilon$ (Fig.~\ref{fig:4}(c)), with the average showing a mild dependence on $\epsilon$ (inset). 

Interestingly the values of $\langle t_{\rm C}^* \rangle$ are nearly independent of genomic separation (Fig.~\ref{fig:4}(d)). Here too, there is huge variability among different bead pairs, with the inset showing the behaviour when the segment length is averaged over all pairs having the same $s$. However, the interaction strength significantly alters the $\langle t_{\rm C}^*\rangle$ (Fig.~\ref{fig:4}(e)) showing an exponential increase (see sec.~S9 for a discussion on the relation between $t_{\rm C}^*$ and $\epsilon$). This suggests that the $\langle t_{\rm L}^* \rangle$ is determined by the interplay between entropy (resulting from genomic separation) and energy (interaction strength). Once bead pairs come in contact, $\langle t_{\rm C}^* \rangle$ is dominated by the interaction strength. 

For the OFF and SAW states, we also show $\langle t_{\rm L}^* \rangle$ between all pairs of beads as a heatmap (see Fig.~\ref{fig:4}(f) \& (g)). One can quickly note that the range of SAW time scales is much higher than that of the OFF state. This is the consequence of higher $\mu$ for the SAW compared to the OFF state. In the SAW, one can observe that the times are similar for all points having the same distance away from the diagonal (a line parallel to the diagonal axis), suggesting that what matters, in this case, is the inter bead distance ($s$). In contrast, in the OFF state, there is heterogeneity and curvy colour contours suggesting that the time values are not just a function of segment length alone but also the identity (interaction strength) of the individual bead pairs. In other words, If entropy is the only thing that mattered, the time values will depend only on the genomic distance (like in the no-interaction SAW case). But here, as observed in Fig.~\ref{fig:4}(c), the interaction strength values also play a role, albeit small compared to the role of entropy. Hence the curvy colour contours in Fig.~\ref{fig:4}(f) once again points to the interplay between energy and entropy. 

\subsection{Nature of loop formation and contact time distributions}
\begin{figure*}[hbt!] 
	\begin{center}
			{\includegraphics*[width=0.85\linewidth]{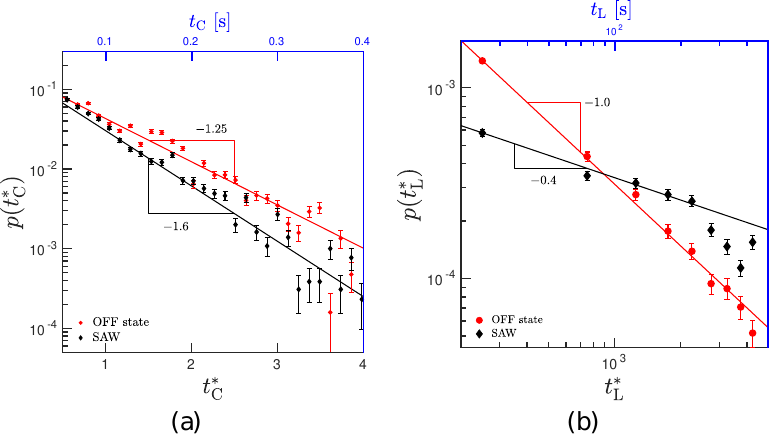}}
		\caption{Distribution function for contact time ($t_{\rm C}^*$) and loop formation time ($t_{\rm L}^*$) for a specific bead-pair (bead 5 and bead 30) in SAW and OFF state are shown in (a) and (b), respectively. \label{fig:5} }
	\end{center}
\vspace{-20pt}	
\end{figure*}

So far, we have studied the average loop formation times and contact times; however, should one assume that the average values describe these quantities completely? To answer this, similar to $p(r^*)$, here we have investigated the nature of the distribution of the temporal quantities. 
In Fig.~\ref{fig:5}(a) and (b), we present the probability distributions of contact ($p(t_{\rm C}^*)$) and loop formation ($p(t_{\rm L}^*)$) times, respectively. We observe that $p(t_{\rm C}^*) \sim \exp{(-t_{\rm C}^*/\tau_{c})}$ with the average time $\tau_c$ that depends on the epigenetic state (SAW: $\tau_{c}= 1/1.6$, OFF: $\tau_{c}= 1/1.25$). $\tau_c$ is small for the SAW and it increases as we add interactions to the system.
However, interestingly, the probability of loop formation time ($t_{\rm L}^*$) has a power law decay ($p(t_{\rm L}^*) \sim {(t_{\rm L}^*)^{-\gamma}}$) (Also see Fig.~S16). This suggests that there is a huge diversity in loop formation times, and the average looping time alone may not be sufficient to describe the loop formation phenomena. We find that the epigenetic states alter the slope of the distribution (SAW: $\gamma= 0.4$, OFF: $\gamma=1.0$), keeping the overall nature the same.
Comparison of these two distributions reveals that quantitatively the $t_{\rm L}^*$ is much larger than $\tau_{c}^*$, indicating that chromatin segments take longer to come into contact but stay in contact for a short time. Interestingly, earlier studies for yeast and drosophila have reported broadly comparable behaviour~\cite{ghosh2018epigenome}. Also, see Fig.~S16 for a discussion on finite size effects introducing exponential tails in the distribution. 

\subsection{Suggestions for experiments to test our predictions} 
Since we take HiC-like data as input and predict positional fluctuations and dynamics of chromatin segments, microscopy is the ideal method to test our predictions~\cite{germier2017real,bintu2018super,nozaki2017dynamic}. All $p(r)$ predictions (Fig.~\ref{fig:1}) may be tested either via live (without fixing) microscopy experiments or by collecting a large number of frozen snapshots of segment-locations via FISH or equivalent methods. Imaging experiments may also estimate the volume occupied by a domain (Fig.~\ref{fig:rg2}). From the positional fluctuation data, one can also obtain the effective stiffness as described earlier in this paper (Fig.~\ref{fig:3}). To measure the time-dependent quantities (Fig.~\ref{fig:4}, Fig.~\ref{fig:5}), apart from live microscopy experiments, one may also design appropriate FRET pairs that can probe quantities like the contact time~\cite{yang2006chromatin}. Obtaining all of these quantities for different epigenetic states would facilitate comparison with our predictions.

\section{Discussion}
In this paper, we have investigated the dynamics of chromatin domains, using an inverse Brownian dynamics model including hydrodynamic interactions, and presented the corresponding results. In this section, we discuss several aspects of the model and results in greater detail. In particular, we would like to highlight the novel aspects of our model in the context of earlier work in this area. 
In our work, we focus on chromatin organization and dynamics within a domain, using parameters obtained via inverse optimization, consistent with chromatin conformation capture experiments. 

In the literature, there have been many remarkable studies to simulate chromatin organization within the nucleus. A number of models have used the forward method to obtain 3D configurations and dynamics of chromatin~\cite{sandholtz2020physical,macpherson2018bottom,fudenberg2016formation,brackley2016predicting,jost2014modeling}. In the forward method, authors start with a fixed set of parameters and the system is simulated to obtain time-dependent/steady-state configurations, typically using Monte Carlo or coarse-grained Molecular dynamics methods. Such studies have been used to obtain 3D organization of yeast genome, regions of drosophila, mouse and human genome~\cite{jost2014modeling,liu2018chain,shukron2017transient,di2018anomalous}. 
Many of these studies point out interesting statistical properties of the chromatin polymer, such as the scaling of contact probabilities with genomic separation and root mean square spatial distance. They also investigate the organization of the active/inactive regions within a chromatin/within the nucleus~\cite{sandholtz2020physical,macpherson2018bottom,fudenberg2016formation,liu2018chain,saintillan2018extensile}. It may be noted that chromatin folding models have broad similarity with protein-folding models  with the interchromatin interactions being the non-bonded interaction~\cite{ueda1978studies,yadahalli2014modeling,gosavi2008extracting,cheung2005molecular,vskrbic2012role,di2016hi}. 

There have been a few models that compute chromatin organization starting with the Hi-C data. Some of these models assume a mathematical relation between the contact probability and mean 3D distance and covert the contact matrix to a distance matrix. A subgroup of these models employ a harmonic interaction between various bead pairs and find optimal spring constants, leading to a Gaussian model~\cite{shi2021hi,shukron2017transient,shinkai2020phi}. Another approach has been to predict a population of static configurations by iteratively inserting contacts and finding the optimal structures that maximize a likelihood function connecting the contact matrix and the model~\cite{hua2018producing,tjong2016population}. In a related approach, groups have investigated the problem using a general force-field optimization method~\cite{qi2019predicting,DiPierro2016}. A different approach has been to employ a coarse-grained polymer model explicitly accounting for binder proteins; these models obtain the optimal number of protein types and binding sites that matched with Hi-C data~\cite{conte2020polymer}. While these models have the details about the protein binding locations, they do not optimize for the interaction strengths directly; once they obtain the optimal number of protein types and bindings sites, the concentration and interaction strengths of proteins are varied as separate parameters. An alternative approach has been to start with the epigenetic (histone modification/protein binding) data and compute interaction strengths that can generate 3D configurations~\cite{DiPierro2016,liu2018chain}.

While these models provide many exciting results and predictions, another approach is to optimise for short-range interaction potentials representing ``bonds" that can be broken when segments pairs are far away. Giorgetti et al. and our own work have attempted to do this~\cite{giorgetti2014predictive,tiana2016structural,kumari2020computing}. While Giorgetti et al. do Monte Carlo simulation of a region of mouse X-chromosome to compute the 3D distance between segments, our focus is to study a wide range of static and dynamic properties of chromatin domains in human chromosomes, using Brownian dynamics, with hydrodynamic interactions. 
Furthermore, Giorgetti et al. use a square well potential; we use the SDK potential that has attractive and repulsive parts that smoothly vary along $r$, reaching zero at $r_c$. There have been a couple of interesting polymer simulation studies on $\alpha$ globin gene locus~\cite{brackley2016predicting,chiariello2020dynamic}. Both the studies investigate the mouse $\alpha$ globin domain, which appears to have a very different contact matrix from the human $\alpha$-globin that we investigate here. Our studies are complementary to these studies as we predict the interaction strengths, 3D distance distributions and dynamics of human $\alpha$ globin locus that have not been investigated so far.

To study dynamics, there have been many interesting studies that compute mean squared displacement (MSD) of a loci as a function of time and showed whether regions have diffusive or sub-diffusive nature depending on the organism, nature of interactions, and length scale~\cite{tortora2020chromosome,ghosh2018epigenome,zhang2016first,khanna2019chromosome,shi2018interphase,di2018anomalous,shinkai2020phi,shukron2017transient}. In the context of yeast and regions of drosophila, mouse, and the human genome, there have been investigations of first meeting times and contact times~\cite{ghosh2018epigenome,zhang2016first,khanna2019chromosome}. Here, we also compute similar quantities and show how the single chromatin domain behaves, taking two loci as examples, with optimised parameters and what the corresponding timescales are. While the precise timescale would vary depending on the nature of the interaction/domain, some features like the power law distributions are common to the earlier observations in drosophila \cite{ghosh2018epigenome}. 

Since many methods use epigenetic datasets to model chromatin, we would like to argue that using Hi-C data alone has the following advantage. In many a context --- for example, when studies are done under different drug-treated conditions --- we may not have the epigenetic data. We may not know the precise protein concentrations, affinities of proteins, organization of proteins and histone marks under all these conditions. Hence obtaining optimal intra-chromatin interaction strengths for all bead pairs with Hi-C data alone has its advantage and may be useful to study various cell states. In this work, given the contact probability data -- without any further information -- we predict the intra-chromatin interactions that are a net result of different epigenetic marks, their intensities, different protein/DNA interactions and distributions/concentrations of proteins.

Simulating chromatin at higher resolution is computationally expensive, in particular with HI. Hence we chose 10kb as the optimal resolution for this work. While the chromatin polymer at 10kb resolution can capture the size, volume, shape and overall dynamics of typical TADs, it cannot capture the enhancer-promoter looping and gene regulatory dynamics. 

Another point worthy of discussion is whether considering a small domain is justifiable or not. First of all, it is computationally expensive to simulate the whole chromatin at higher resolution with hydrodynamic interactions; moreover, from a biology point of view, there are sound arguments that suggest that simulating a small domain is reasonable. From the chromatin contact probability data, we know that chromatin is organised into multiple domains (also often called TADs). This data suggest that one domain has much less probability (negligible probability) to interact with another domain. Hence their inter-domain interaction can be assumed to be negligible. Since our aim here is to probe the dynamics of beads within a domain, it is reasonable to simulate a single domain alone. Also, note that the gene regulation and cellular processes are thought to be happening within a domain. Interestingly, recent experimental data has shown that when the chromatin polymer is cut and each domain is separated out, most of those domains are stable and behave very similar to how they behaved within the whole chromatin, indicating that each domain is independent~\cite{belaghzal2021liquid}. This justifies our investigation of each chromatin domain as a separate entity. Interestingly, the intra-chromatin contacts from HiC will shrink the domain and make it dense. We find that the volumic density within the domain — defined as the ratio of the segment length to the volume of the domain — is 0.005bp/nm$^3$. This is the same order of volumic density as mentioned in~\citet{ghosh2018epigenome}.

One of the important conclusions of polymer physics is that the temporal quantities are greatly affected by HI~\cite{pham2008brownian,pham2010collapse,prakash2009micro,prakash2019universal,prabhakar2017effect,prabhakar2004multiplicative,schroeder2018single,sunthar2006dynamic,sunthar2005parameter,schroeder2004effect}.
 Earlier works that used HI investigated the active dynamics of a long polymer with the aim of investigating correlated motion due to activity~\cite{saintillan2018extensile}. While these studies gave insights about the potential large scale movements inside the nucleus due to activity, the dynamics of chromatin in the length scale relevant for gene regulation (scale of domain/TADs) is not investigated here.
To the best of our knowledge, this work is the first attempt to study the chromatin dynamics with HI at the length scale of a domain. 
Here, we show that while the relaxation times do depend on HI, other quantities on the length scale of the domain remain less affected by HI. However, there are interesting questions that remain to be addressed on how HI affects dynamics in a crowded environment, given that the concentration of macromolecules in a cell is in an unentangled semi-dilute regime, where HI is known to be important. However, investigating dynamics in a crowded environment is a difficult problem given that HI is a long-ranged and involve many-body interactions. The problem of the role of screening of HI is also important in this context. Our current study is a preliminary step in the direction of including HI. Another aspect worthy of discussion is alternative models like the melt model~\cite{bohn2010topological,rosa2014ring,halverson2014melt}. The melt of polymer rings could be a useful model to study organization of multiple chromosomes and how chromosomes segregate to form chromosome territories~\cite{bohn2010topological,rosa2014ring,halverson2014melt}. It could also be an interesting model to study how multiple TADs do segregate. However, in the current work, we are neither studying the problem in the length scale of multiple chromosomes nor multiple TADs. Rather, we are interested in the chromosome structure/dynamics within a TAD/domain and the use of a model with a single chain in a dilute solution capable of providing insights in this context.

Quantities calculated here have immense physical and biological significance. The finding of two peaks in the distribution function is a novel aspect that can have ramifications.
As mentioned earlier, there is an ongoing debate in the field about whether gene regulation requires actual physical contact between two regulatory segments or only the proximity would suffice. Cellular processes such as transport of proteins from one region to another (e.g. enhancer-promoter), spreading of histone modifications in the 3D space etc., would crucially depend on $p(r)$~\cite{katava2021chromatin,jost2018epigenomics}. For example, given $r$, one can compute the time ($\tau_p$) for proteins/enzymes to diffuse from location $r_{i}$ to $r_{j}$. The mean time would depend on the distribution as $\langle \tau_p \rangle = \int \tau_p p(r) d r$. However, apart from a distance among segments, the accessibility would depend on the local compactness and diffusivity too. That is, compactness of the domain (Fig.~\ref{fig:rg2}) and effective viscous drag (Fig.~\ref{fig:3}) together with $p(r)$(Fig.~\ref{fig:1}) would be crucial for understanding how physics of chromatin would affect biological function. Given that phase separation is argued to be one of the important factors determining domain formation, our study reveals how the interplay between intra-chromatin interactions and polymer dynamics would affect loop formation and contact times. 

\section{Conclusion}
Even though there is a great improvement in our understanding of the static nature of chromatin organization, very little is known about the dynamics, which is a crucial aspect of \emph{in vivo} chromatin. 
With the advancement of technology, it is now possible to experimentally probe the fluctuations and dynamics of chromatin polymer. However, the main challenge to simulate the dynamics of chromatin is that we do not know the interaction strength parameters among different segments. We have overcome this challenge by using an inverse technique and obtained optimal interaction strengths between all chromatin segments, and used it to investigate the dynamics of a chromatin domain. 

We summarize our key findings: 
(i) We investigated the 3D organization of two chromatin domains for two different epigenetic states in each case. (ii) Starting from 5C/HiC data, we predicted the optimal intra-chromatin interactions strengths for all cases revealing how epigenetic changes would affect the interactions and 3D organization of chromatin. 
(iii) Going beyond the average properties, we computed the distance probability distribution; we observe a double peak -- an interaction-driven peak and an entropy-dominated peak -- depending on the epigenetic state and chromatin segment pair identity.
(iv) Introducing perturbations in optimized interaction strength values to systematically mimic epigenetic-like states, we show how perturbations would alter $p(r)$; the distance distribution between a given bead pair depends on the interaction strength of all other pairs suggesting the cooperative nature of chromatin folding. 
(v) Volume and the shape properties of the chromatin domain depends on the magnitude of interaction strengths present, whether it is an epigenetic state or perturbed state. The OFF state of $\alpha$-globin gene is highly collapsed/compact, more spherical compared to the extended, less spherical ON state or SAW.
(vi) Our simulations investigated beyond the pairwise contact probability information and predicted the probability of three segments coming together (triple contact).
(vii) The relaxation time of the domain is dependent on the magnitude of the interaction strengths in the domain. Counter-intuitively, the relaxation time of a highly interacting OFF state is much shorter than that of a non-interacting SAW polymer. We explain this phenomenon by computing the effective stiffness of the domain from polymer fluctuations. We also show that the OFF state has a higher effective drag.
(viii) We study dynamics accounting for crucial hydrodynamic interactions; we show that HI has a significant influence on the relaxation time of the chromatin domain. With HI, the domain takes half the time to relax as compared to the no-HI case.
(ix) We compute the loop formation time and the time for the looped bead pairs to remain in contact. We show that average looping time has different scaling with genomic separation, depending on the nature of the chromatin states having different interaction strengths. The looping times show a power law distribution indicating multiple timescales that might be involved with looping. On the other hand, the contact time has an exponential distribution. 

This study can be further extended genomewide to examine various gene loci and investigate the fluctuations and dynamics of all domains in the genome. Such polymer models are useful for examining aspects like the spread of histone modifications and accessibility of the domains. We hope that this study will catalyse new experimental and computational studies examining the interplay between epigenetics and polymer dynamics. 

\noindent
\textbf{Acknowledgments}
We thank Burkhard D\"unweg, Dibyendu Das and Rajarshi Chakraborty for enlightening discussions. The work was supported by the MonARCH and SpaceTime computational facilities of Monash University and IIT Bombay, respectively. We also acknowledge the funding and general support received from the IITB-Monash Research Academy, DST, SERB and DBT India.

\noindent
\textbf{Conflict of Interests:} The authors declare that they have no conflict of interest.

\end{document}

\title{ {\huge Supplementary Information}\\
Heterogeneous interactions and polymer entropy decide organization and dynamics of chromatin domains}

\author{Kiran Kumari}
\affiliation{IITB-Monash Research Academy, Indian Institute of Technology Bombay, Mumbai, Maharashtra -  400076, India}
\affiliation{Department of Biosciences and Bioengineering, Indian Institute of Technology Bombay, Mumbai 400076, India}
\affiliation{Department of Chemical Engineering, Monash University, Melbourne, VIC 3800, Australia}
\author{J. Ravi Prakash}
\affiliation{Department of Chemical Engineering, Monash University, Melbourne, VIC 3800, Australia}
\author{Ranjith Padinhateeri}
\affiliation{Department of Biosciences and Bioengineering, Indian Institute of Technology Bombay, Mumbai 400076, India}

\maketitle
	
	The main paper presents findings of studies on statics and dynamics of a chromatin domain. Through chromatin {\it in vivo} shows dynamicity and cell-to-cell variability; in literature, the nature of chromatin is mostly quantified through average static properties. In this work, we go beyond the average and quantify the whole phase space by investigating probability distributions and several temporal quantities, and we estimate the timescales of loop formation and stable loop maintenance. The necessary supplementary information for results presented in the main paper is provided here. 
	
%	This document is organised as follows : Sec.~\ref{sec:simpara} describes the simulation technique, governing equations and Inverse Brownian dynamics (IBD) algorithm. The procedure to convert the non-dimensional quantities obtained from simulation to its corresponding standard experimentally measured units is discussed in Sec.~\ref{sec:units}.  We perturb the interaction strengths systematically to model different epigenetic states as described in Sec.~\ref{sec:epigen}. Validation of distance probability distribution with SAW and cumulative distribution for $\alpha$-globin gene is presented in Sec.~\ref{sec:pr}. The shape analysis of $\alpha$-globin domain is given in Sec.~\ref{sec:shape}. Finally, Sec.~\ref{sec:temporal} contains the analysis of time-dependent quantities. 

\section{\label{sec:simpara}Simulation details}
 We consider chromatin as a bead spring chain having optimal intra-chromatin interactions derived from 5C and Hi-C data using an inverse method described below. 
Taking experimental contact probability data as input, we have computed the optimal interaction strength between different segments of the chromatin using an Inverse Brownian Dynamics (IBD) method that we have developed~\cite{kumari2020computing}. 
We perform Brownian dynamics simulations and compute various static and dynamic quantities as described below. 
It is essential to include hydrodynamic interactions (HI) while computing dynamic quantities in order to obtain even qualitatively accurate predictions~\cite{pham2008brownian,pham2010collapse,prakash2009micro,prakash2019universal,prabhakar2017effect,prabhakar2004multiplicative,schroeder2018single,sunthar2006dynamic,sunthar2005parameter,schroeder2004effect}. 
HI accounts for the propagation of velocity perturbations due to the motion of one segment of a polymer chain to other segments through the medium. Several studies that attempt to probe chromatin dynamics have not considered HI. In the present work, we simulate the dynamic nature of chromatin accounting for HI.

\subsection*{Governing equations for the bead-spring chain model}
The total energy of the chromatin bead spring chain, made up of $N$ beads, is $U=U^{\rm S} +U^{\rm SDK}$ where $U^{\rm S}$ is the spring potential between the adjacent beads $i$ and $(i+1)$, given by 
\begin{align}
U^{\rm S} =\sum_{i} \frac{H}{2}(|\bm r_{i}-\bm r_{i+1}| - r_0)^2
\end{align}
where $\bm r_{i}$ is the position vector of bead $i$, $r_0$ is the natural length and $H$ is the stiffness of the spring. 
The Soddemann-Duenweg-Kremer potential ($U^{\rm SDK}$) is a Lennard-Jones-like potential used to mimic protein-mediated interactions~\cite{soddemann2001generic,santra2019universality} which is given by:

\begin{align}\label{eq:SDK}
\small
U^{\textrm{SDK}}=\left\{
\begin{array}{l l l}
			&  4\left[ \left( \dfrac{\sigma}{r_{ij}} \right)^{12} - \left(\dfrac{\sigma}{r_{ij}} \right)^6 + \frac{1}{4} \right] - \epsilon_{ ij}  ~& r_{ij}\leq 2^{\frac{1}{6}}\sigma \\
			& \frac{1}{2} \epsilon_{ij} \left[ \cos \,(\alpha r_{ij}^2+ \beta) - 1 \right] & 2^{\frac{1}{6}}\sigma \leq r_{ij} \leq r_c \\
			& 0 &  r_{ij} \geq r_c.
\end{array}\right.
\end{align}
Here $ r_{ij}=|\bm r_{i}-\bm r_{j}|$ is the distance between beads $i$ and $j$ and $\epsilon_{ij}$ is an independent parameter representing the attractive interaction strength between them. $2^{1/6}\sigma$ is the distance at which $U^{\rm SDK}$ is zero. 
The cut-off radius of the SDK potential is $ r_c = 1.82\sigma$.
The SDK potential has the following advantages compared to the Lennard-Jones (LJ) potential: 
\begin{enumerate}
\item The repulsive part of the SDK potential ($ r_{ij}\leq 2^{1/6}\sigma$) is constant and remains unaffected by the choice of the parameter $\epsilon_{ij}$. 
\item The SDK potential reaches zero at the cut off radius $r_c = 1.82\sigma$, unlike the LJ potential where the energy goes to zero only at infinite distance. 
\item As the protein-mediated crosslinking interactions in chromatin are short ranged, the SDK potential models the chromatin crosslinks more appropriately~\cite{soddemann2001generic,Steinhauser2005a}. 
\end{enumerate}

For the simulation, all the length and time scales are non-dimensionalised with $l_H=\sqrt{k_BT/H}$ and $\lambda_H=\zeta/4H$, respectively where $T$ is the absolute temperature, $k_B$ is the Boltzmann constant, and $\zeta=6\pi\eta_s a$ is the Stokes friction coefficient of a spherical bead of radius $a$, with $\eta_s$ being the solvent viscosity. The evolution of bead positions in BD simulations is governed by the following Ito stochastic differential equation,
\begin{align}\label{eq:sde_eq}
\begin{aligned}
\bm r_{i}^*(t^* + \Delta t^*) = \bm r_{i}^*(t^*)  + &\frac{\Delta t^*}{4}   \bm D_{ij}\cdot (\bm F_{j}^{S*}+ \bm F_{j}^{\rm SDK*}) \\+&\frac{1}{\sqrt{2}}\bm B_{ij} \cdot \Delta \bm  W_{j}
\end{aligned}
\end{align}
Here $t^* = t/\lambda_H$ is the dimensionless time and $r_{i}^* = r_{i}/l_H$ is the dimensionless length.   
$\Delta\pmb W_{j}$ is a non-dimensional Wiener process with mean zero and variance $\Delta t^*$.
The bonded interactions between the beads are represented by a non-dimensional spring force, $\mathbf F_{j}^{\rm S*}$, and the non-dimensional SDK force is $\mathbf F_{ j}^{\textrm{SDK*}}$.
 $\pmb D_{ij}$ is the diffusion tensor, defined as $\pmb D_{ij} = \delta_{ij} \pmb \delta + \pmb \Omega_{ij}$, where $\delta_{ij}$ is the Kronecker delta, $\pmb \delta$ is the unit tensor, and $\pmb{\Omega}_{ij}$ is the hydrodynamic interaction tensor.  We use the regularized Rotne-Prager-Yamakawa (RPY) tensor to compute hydrodynamic interactions (HI)~\cite{prakash2019universal},
\begin{align}
{\bf \Omega}(\mathbf{r^*}) =  {\Omega_1{ \boldsymbol \delta} +\Omega_2\frac{\bf{r^*}{r^*}}{{r^*}^2}}
\end{align}
for $r^*\geq 2\sqrt{\pi}h^*$
\begin{align}
\begin{aligned}
\Omega_1 =& \frac{3\sqrt{\pi}}{4} \frac{h^*}{r^*}\left({1+\frac{2\pi}{3}\frac{{h^*}^2}{{r^*}^2}}\right);\\
\Omega_2 =& \frac{3\sqrt{\pi}}{4} \frac{h^*}{r^*} \left({1-\frac{2\pi}{3}\frac{{h^*}^2}{{r^*}^2}}\right) 
\end{aligned}
\end{align}
and for $ r^*\leq 2\sqrt{\pi}h^*$
\begin{align}
\begin{aligned}
\Omega_1 =& 1- \frac{9}{32} \frac{r^*}{h^*\sqrt{\pi}};\\
\Omega_2 =& \frac{3}{32} \frac{r^*}{h^*\sqrt{\pi}} 
\end{aligned}
\end{align}
Here $h^*$ is the dimensionless hydrodynamic parameter, defined as $h^* = a/(l_{H}\sqrt{\pi})$. In the present work, we set $h^*=0.25$~\cite{sunthar2005parameter}. 
The range of values for $h^*$ is fairly narrow ($0 \leq h^* \leq 0.3$) as discussed in~\cite{beris1994thermodynamics}. From polymer physics, we know that the true measure of hydrodynamic interaction is the ``non-draining" parameter $h=h^*\sqrt{N}$, and that properties reach universal value for large values of $N$~\cite{ottinger1987generalized,zimm1956dynamics}. In particular, the value $h^* = 0.25$ is a `fixed-point' value, which leads to faster convergence to universal behaviour, independent of chemical details, and that is why we have used this value~\cite{ottinger1989renormalization}. 
All the simulation parameters with their corresponding values are indicated in Table~\ref{sim_para}.
\begin{table*}[ht]
	\begin{minipage}{\textwidth}
		\setlength{\tabcolsep}{25pt}
		\centering
		\caption{\label{sim_para} Numerical values of all the parameters used in the simulation }
		\begin{center}
			\begin{tabular}{c c c}
				\hline  \hline Simulation parameters & symbols & values \\
				\hline
				no. of beads 		&	 $N$           &    50   \\
				cut-off radius		& $r_{c}^*$                &    $1.82~\sigma^*$  \\
				natural length of the spring &$r_{0}^*$                &    $1~\sigma^*$  \\ 
				$U^{\rm SDK}$ length scale 	& $\sigma^*$            	&    1    \\
				hydrodynamic parameter 	& $h^*$            	&    0.25    \\ 
				SDK parameter		&$\alpha$      	     &    1.53063    \\ 
				SDK parameter		&$\beta$         &    1.21311    \\ 
				%\hline
				 length scale		&$l_H$         &    36 nm    \\ 
				 time scale	&$\lambda_H$         &    0.1 s    \\ 
				 interaction strength	&$\epsilon_{ij}$         &   from IBD   \\ 				
				\hline \hline
			\end{tabular}
		\end{center}
	\end{minipage}
\end{table*} 

\subsubsection*{\label{sec:ibd} Inverse Brownian Dynamics (IBD)}
\begin{figure}[hbt!] 
		\centering
	{\includegraphics*[width=3in,height=!]{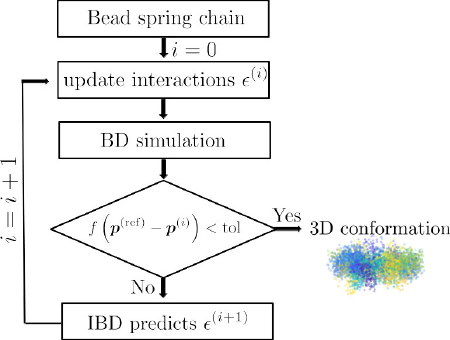}} 
\caption{A flowchart representation of the inverse Brownian dynamics algorithm. Each iteration consists of BD simulation, calculation of contact probability and revision of interaction strength. Here, $p^{\rm (ref)}$ represents the reference contact probability matrix, and $p^{\rm (i)}$ represents the contact probability matrix from simulations at iteration i. This figure is reproduced from Kumari et al.~\cite{kumari2020computing}. \label{fig:S1}}
\end{figure}

To simulate chromatin dynamics accurately, one requires the potential energy parameters that specify the interaction strengths ($\epsilon_{ij}$) between different regions of chromatin. We have developed an IBD algorithm that extracts the optimal interaction strengths between bead pairs from an experimental contact probability map~\cite{kumari2020computing}. The details of the algorithm are given below:

The essence of the algorithm is to find the average contact probability and iteratively compare it to the reference contact probability obtained from experiments. 
The average contact probability $ p_{m}$ of the bead-pair $m$ is given by 
\begin{align}\label{eq.probab}
p_{m}= \langle \hat{p}_{m}\rangle = \dfrac{1}{Z}\int d\Gamma \, \hat{p}_{m} \exp (-\beta \mathcal{H})
\end{align}
Here $Z=\int d\Gamma \exp (-\beta \mathcal{H})$ is the partition function and $\hat{p}_{m}$ is an indicator function which indicates when contact occurs between the bead pair represented by index $m$. 
$\hat{p}_{m}$ is 1 if the distance between the beads is less than the cut-off distance of the indicator function, $r_p^*$, and 0 otherwise. For this work $r_p^*$ = $r_c^*$ = $1.82 \sigma$ is taken as the cut-off distance/capture radius of the SDK potential. For our studies $\sigma$ is of the order of $36$nm($1.82 \sigma \approx 65$nm).
 We do not know the value of the capture radius ($r_p^*$) precisely. However, \citet{giorgetti2014predictive}, have previously systematically varied a similar capture radius parameter and have reported a value of roughly the same order as that used here (between 50nm and 100nm). 

In a separate study, we have used the SDK potential to systematically study the behaviour of polymers in a poor solvent, with the capture radius as a parameter, and have reproduced the standard scaling in polymer physics accurately. It was found that $r_c =1.82\sigma$ was the optimum value for reproducing homogenous polymer behaviour in poor solvents~\cite{santra2019universality}. \\

As the interaction strengths are defined between a pair of beads, we use a single index ($m$) to define a specific bead pair ($i$ and $j$) as $m = \frac{1}{2}[ i(i -1) ] + [j - (i -1)] $. Here $i$ varies from $2$ to $N$, and $j$ varies from $1$ to $(i -1)$ for a matrix of size $N$. $\hat{p}_{m}$ is 1 if the distance between the beads is less than the cut-off distance of the indicator function, $r_{\textrm{c}}^*$, and $0$ otherwise. We intend to target the experimentally obtained contact probability $p_{m}^{\rm ref}$ by adjusting the well-depth of SDK attractive interactions $\epsilon_{m}$. The Taylor series expansion of $\langle\hat{p}_{m}\rangle $ about the interaction strength $ \epsilon_m$ after neglecting higher order terms is 
\begin{align}\label{eq.ibd_main}
\langle \hat{p}_{m}\rangle(\epsilon_{m} + \Delta \epsilon_{m}) = \left\langle \hat{p}_{m}\right\rangle(\epsilon_{m}) +  \sum_n \chi_{mn}\,\Delta \epsilon_n 
\end{align}
where $\Delta \epsilon_{m}$ is the change in the interaction strength, and the susceptibility matrix 
\begin{align}
\chi_{mn} = \dfrac{\partial \langle \hat{p}_{m}\rangle}{\partial \epsilon_n} =  \dfrac{\partial}{\partial \epsilon_n}\left[\dfrac{1}{Z}\int d\Gamma \hat{p}_{m} \exp (-\beta \mathcal{H})\right]
\end{align}
\begin{align}
\begin{aligned}
\chi_{mn}= &\beta\left[\frac{1}{Z}   \int \hat{p}_{m}  b_n\exp(-\beta\mathcal{H})d\Gamma\right.\\ 
-&\left. \langle \hat{p}_{m}\rangle\frac{1}{Z}\int b_n\exp (-\beta \mathcal{H})d\Gamma \right] 
\end{aligned}
\end{align}
\begin{align*}
\chi_{mn}= \beta\left[ \langle \hat{p}_{m} b_n\rangle -\langle \hat{p}_{m}\rangle \langle b_n\rangle\right]
\end{align*}
where
\begin{align}
b_n = -\frac{\partial \mathcal{H}}{\partial \epsilon_n} 
\end{align}
Replacing the left hand side of Eq.~\ref{eq.ibd_main} with the target contact probability $p_m^{\rm ref}$ obtained from experiment, we get 
\begin{align}\label{ibd}
p_m^{\rm ref} - \langle \hat{p}_{m}\rangle = \sum_n \chi_{mn} \, \Delta\epsilon_n
\end{align}
Equation~\ref{ibd} can be solved for any particular iteration step as 
\begin{align}
\epsilon_n^{ (i+1)}=\epsilon_n^{ (i)}+ \lambda \sum_m { C}_{nm}^{ (i)}\left(p^{\rm ref} - \langle \hat{p}_{m}\rangle^{ (i)}\right)
\end{align}
where the matrix $\mathsf{C}$ is the $pseudo$-$inverse$ of the matrix $\mathsf{\chi}$~\cite{kumari2020computing}, superscript $i$ represents the iteration number, $\lambda$ denotes the damping factor with $0<\lambda<1$, and $\epsilon_n^{(i+1)}$ is the well-depth of the SDK attractive interaction for the next iteration step. Note that when we iteratively tune the interaction strength, we are tuning for the whole set of bead-pairs simultaneously. The parameter ``susceptibility" in the equation accounts for the bead interaction with all other beads. 

The flowchart of the IBD methodology is given schematically in Fig.~\ref{fig:S1}. We start with the initial guess values of interaction strengths, simulate the polymer following the conventional forward Brownian dynamics method and obtain the simulated contact probabilities in the steady state.
The interaction strengths are revised for the next iteration, depending upon the difference between the simulated and the known experimental contact probabilities.
We perform several iterations of the loop (i.e., BD simulation, calculation of contact probability, revision of interaction strength) until the error between the simulated and experimental contact probabilities is less than a predetermined tolerance value. Using these optimal interaction strengths, we study the dynamics of the chromatin domain. In the present work, we have applied the IBD technique to the two chromatin domains ($\alpha$-globin gene locus and a domain in Chr7) and investigate the static and dynamic properties of the chromatin domain. 

\begin{figure*}[hbt!] 
		\centering
	{\includegraphics*[width=0.8\linewidth]{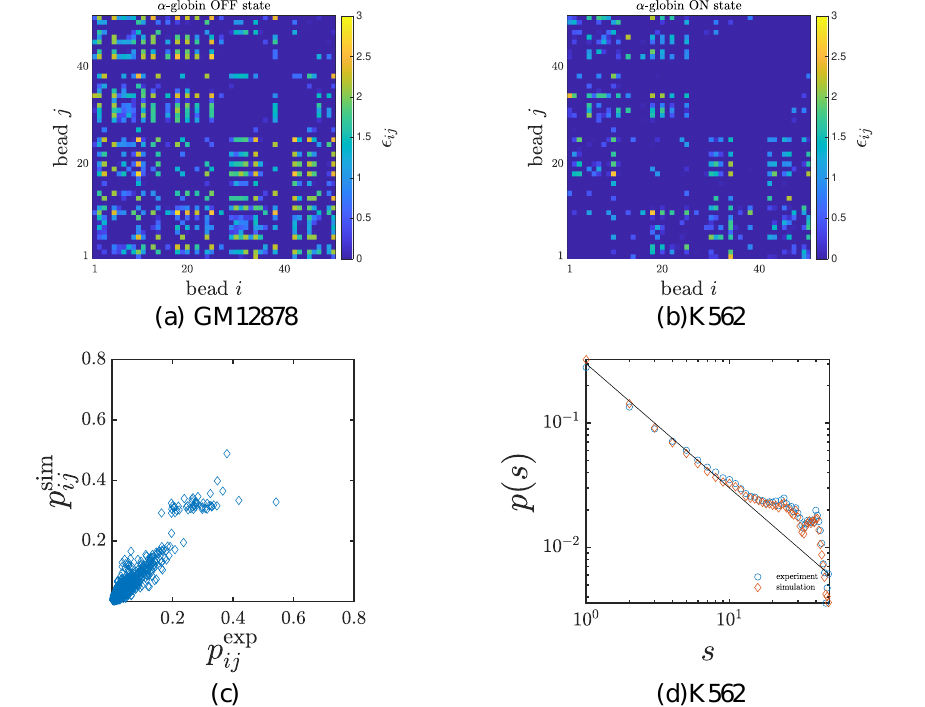}} 
\caption{(a) and (b) show the optimised interaction strength for OFF (GM12878) and ON (K562) state of $\alpha$-globin, respectively. (c) Correlation between the experimental and simulated contact probability. The Pearson correlation coefficient is 0.94. (d) Contact probability as a function of inter bead distance from experiment and simulation. The curve $p(s) \propto s^{-1}$ is shown as a guide to the eye. 
\label{SI:optimised-eps-ag}
}
\end{figure*}

The error between the reference contact map and the contact map computed from simulations decreases and converges  to a small value, smaller than a desired tolerance value.
The optimised interaction strength ($\epsilon_{ij}$) values for $\alpha$-globin GM12878 and K562 are shown in the Fig.~\ref{SI:optimised-eps-ag}(a) and (b),respectively. Fig.~\ref{SI:optimised-eps-ag}(c) shows the correlation between the experimental and simulated contact probability for K562 cell line. The Pearson correlation coefficient is 0.94. Contact probability as a function of inter bead distance from experiment and simulation is shown in Fig.~\ref{SI:optimised-eps-ag}(d).

%\section{Optimised interaction strength for $\alpha$-globin gene from IBD}

\section{Equation predicting interaction strength ($\epsilon_{ij}$) from contact probability ($p_{ij}$) \label{sec:SI_forumla}}
\begin{figure*}[hbt!] 
		\centering
	{\includegraphics*[width=0.8\linewidth]{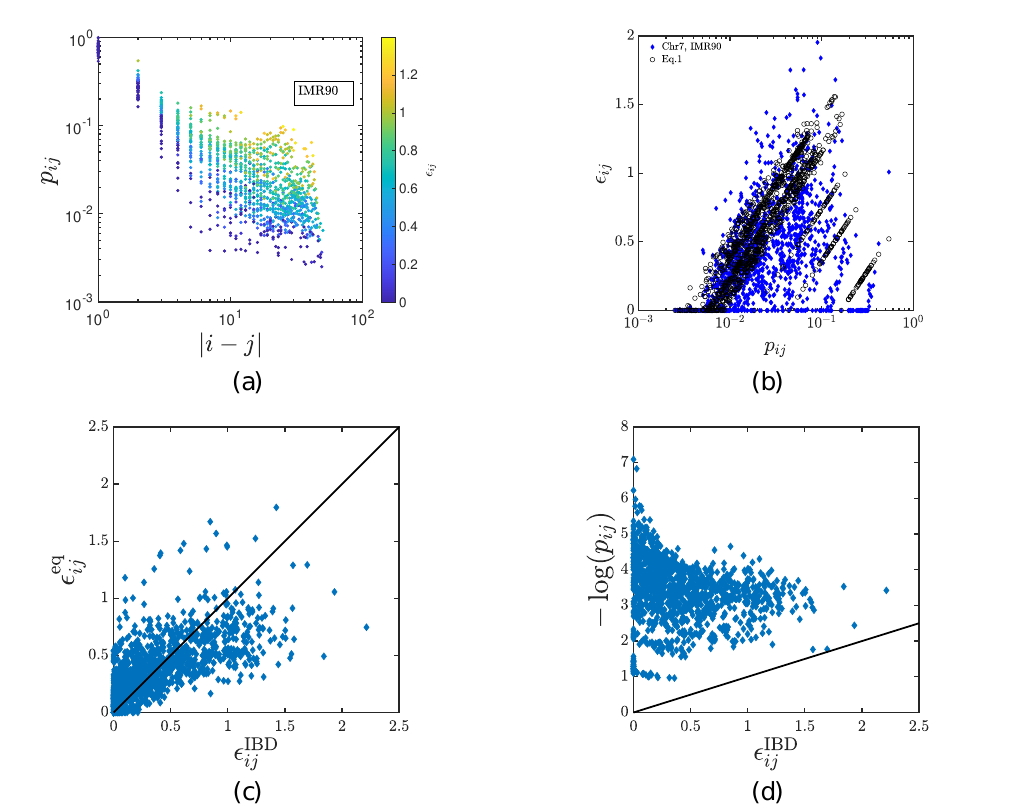}} 
\caption{(a) Our prediction of the relation between interaction strengths ($\epsilon_{ij}$), contact probability and segment length ($|i-j|$) from IBD simulations. See colorbar for $\epsilon_{ij}$. (b) Prediction of the relation between interaction strength and contact probability from the IBD simulation (filled diamonds) and Eq.1(black open circles) of the main manuscript for the Chr.7 region from IMR90 cell line.
(c) Scatter plot of $\epsilon_{ij}$ from IBD in the $x$-axis and $\epsilon_{ij}$ from the equation (Eq.~1 in the main text) in the $y$-axis.  There is a good positive correlation between the two, and the points fall near the line $y=x$ suggesting that the equation gives a reasonable estimate of the  $\epsilon_{ij}$.  
(d) Scatter plot of $\epsilon_{ij}$ from IBD in the $x$-axis and the negative of $\log{p_{ij}}$ in the $y$-axis. $\epsilon_{ij}=-\log{p_{ij}}$ is a naive guess for the interaction strengths. But the points highly deviating from the $y=x$ line suggest that $-\log{p_{ij}}$ is not a good estimate of $\epsilon_{ij}$. Both the plots here are for K562 cell type.
\label{fig:SI_forumla}}
\end{figure*}

To understand how $\epsilon_{ij}$ depends on $p_{ij}$ and the genomic separation ($|i-j|$), we plotted the data from IBD as shown in Fig.~\ref{fig:SI_forumla}(a). This gave us a hint that the $\epsilon_{ij}$ increases by $p_{ij}$ logarithmically ($\epsilon_{ij}$ increases with a factor of 1 while $p_{ij}$ increases by a factor of 10). We also observed that for every $|i-j|$, there exist a minimum value of $p_{ij}$ which we define as $p_{\rm min}(i-j)$. The $p_{\rm min}(i-j)$ for every $|i-j|$ can be easily obtained from the contact probability matrix. This gave us the formula described in the Eq. 1 of the main text. Here, we discuss how well equation. 1 in the main text predicts $\epsilon_{ij}$ values. Fig.~\ref{fig:SI_forumla}(b) shows the relation between interaction strength and contact probability from the IBD simulation and Eq.1 of the main manuscript for the Chr.7 region from the IMR90 cell line. A simple way to test the equation is to have a scatter plot $\epsilon_{ij}$ from IBD in the X-axis and $\epsilon_{ij}$ from the equation (Eq. 1 in the main text) in the Y-axis. as shown in Fig.~\ref{fig:SI_forumla}(c). They fall near the line $y=x$, implying a good correlation between the two. This formula would immensely help in generating excellent initial guess values of $\epsilon_{ij}$ and will lead to faster convergence of the IBD. 
%\tr{Knowing such a formulae would immensely help in a better initial guess and faster convergence of IBD algorithm.}

A naive guess for such a formula would have been $\epsilon_{ij} \propto -\log(p_{ij})$. In Fig.~\ref{fig:SI_forumla}(d), we are testing this formula and found that this simple relation does not hold. 

\section{Scaling of 3D distance with genomic distance: comparison with experiments \label{SI:sec:comp_exp}}
\begin{figure}[hbt!]
	\centering
		{\includegraphics*[width=0.8\linewidth]{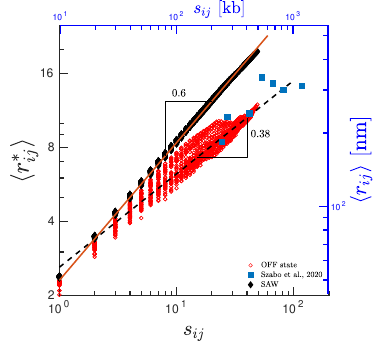}} 
	\caption{ Average 3D distances between all bead-pairs as a function of corresponding genomic distances for the control simulations (SAW, black symbols), chromatin domain that we simulated (red symbols) and comparison with experimental data from ~\cite{szabo2020regulation} (blue symbols).  The major axes (lower $x$ and left $y$) represent quantities in dimensionless units (see methods) while the other axes (upper $x$ and right $y$) represent the same in standard units.   
	\label{fig:comp_exp}}
\end{figure} 
We computed the spatial distance in the repressed state of human $\alpha$-globin gene (OFF state) and found $\langle r_{ij}^* \rangle \sim s_{ij}^{\nu}$ with a scaling exponent $\nu=0.38$ (Fig.~\ref{fig:comp_exp}). This suggests near close packing within the chromatin domain (red symbols). As a ``control'', we also simulated a self-avoiding walk (SAW) polymer with no attractive interaction ($\epsilon_{ij} =0$) which results in $\nu=0.6$ as expected (black symbols)~\cite{bird1987dynamics}. 
It is interesting that some of the experimentally studied epigenetic state by Zhuang lab \cite{boettiger2016super} shows a scaling of 0.37 which is  close of what we show here. However, considering the heterogeneity of chromatin structure, different epigenetic states in different chromosomes/organism can show different scaling. For instance, Zhuang lab has shown scaling exponent in the range of 0.22 to 0.37 depending on the epigenetic state in Drosophila cells. 
Our chromatin model predicts a spread (variability) in the 3D distance (red symbols) which is absent in the control revealing the implications of heterogeneous intra-chromatin interactions.  
A recent microscopy study~\cite{szabo2020regulation} on a mouse ESC chromatin domain also showed a similar behaviour -- both the scaling (slope) and the variability in the experimental and simulation data are comparable without any fitting parameter. 
This is a validation that macroscopic polymer properties of the chromatin domain in our simulation accurately represent what is observed in realistic systems. The $y$-intercept of the experimental data gives us the size ($\sigma=l_H$) of the 10kb chromatin (a single bead in our simulation). For this experimental system (mouse chromosome 6, $1.2$MB in Szabo et al.~\cite{szabo2020regulation}) we get $l_H=22$nm.  
Even though we do not have such extensive spatial distance data (between all pairs) for human $\alpha$ globin, we used the available FISH data for the distance between a single pair of $\alpha$ globin segments and deduced the $l_H=36$nm. Throughout this paper, $l_H=36$nm and $\lambda_H=0.1$s are used to convert all non-dimensional lengths and times into standard units, and we will present quantities in both units. The reasons for the choice of both these specific values are discussed in greater detail in the next section.

\section{\label{sec:units}Conversion of non-dimensional length and time to standard units }
In our simulations, all quantities are computed in dimensionless units as described earlier. To convert these dimensionless numbers to standard units having appropriate dimensions, we need to determine a length scale and a timescale. By comparing our simulations with appropriate experimental observations, we deduce values of characteristic length and time scales that can be used for the unit conversion as follows:
\begin{figure}[hbt!] 
	\centering
		{\includegraphics*[width=0.8\linewidth]{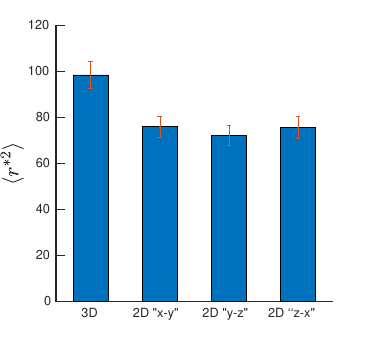}} 
	\caption{Depiction of the calculation of 2D distance between bead $5$ and bead $40$ from simulation.  
	\label{fig:2d}}
\end{figure} 
~\\
{\bf Length scale:} Even though the 3D distances between the genomic segments of $\alpha-$globin are not available, the 2D distance between the two probes located at $34,512$ - $77,058$~bp and $386,139$ - $425,502$~bp for GM12878 (OFF state) was found to be  $318.8 \pm 17.0$ nm from the 2D FISH~\cite{bau2011three}. We computed the average 2D distance ($= 8.81$) between the corresponding bead pair (bead 5 and bead 40) from our simulation by averaging it over all the three 2D planes ($xy$, $yz$, $zx$) as depicted in Fig.~\ref{fig:2d}. By comparing 2D distance values obtained from simulation and experiment, we estimate the characteristic lengthscale in our simulation as $l_H =318.8/8.81 \approx 36$nm. We use this value of $l_H$ to convert all non-dimensional lengths to standard units.    
~\\
{\bf Time scale:} The timescale in our simulation is given by:
\begin{align}
\lambda_H = \frac{\zeta}{4H}=\frac{6\pi \eta_s a^3}{4k_{\rm B}T}
\end{align}
where $H$ is the spring constant, $T$ is the absolute temperature, $k_B$ is the Boltzmann constant, and $\zeta=6\pi\eta_s a$ is the Stokes friction coefficient of a spherical bead of radius $a$ where $\eta_s$ is the solvent viscosity.
For our problem $a= h^*l_H\sqrt{\pi}\approx 16$nm. However, we do not know the precise viscosity of the solvent in the nucleus. There are many estimates ranging over several orders of magnitude from $10^{-3}$~Pa.s to $10^3$~Pa.s. \cite{erdel2015viscoelastic, caragine2018surface}. Given this degree of variability, we decided to use a simple method to estimate time, based on recent experimental reports of chromatin dynamics. Chromatin segments under microscope seems to ``diffuse'' around in a region having the size of the order of $\approx 0.1 (\mu m)^2$ within a timescale of $\approx$ 50 seconds~\cite{germier2017real}. This leads to a diffusion coefficient ($D$) of the order of $500$ nm$^2$/s, and a timescale
\begin{align}
\lambda_H =\frac{a^2}{4D} = \frac{(16~{\rm nm})^2}{4\times 500~{\rm nm}^2/{\rm s}} = 0.12~{\rm s}
\end{align}
Since the calculation is to estimate the order of magnitude number, throughout this work, we use $\lambda_H=0.1$~s. Interestingly, this also corresponds to an effective viscosity roughly in the middle of the wide range estimated previously. These values of $l_H$ and $\lambda_H$ are considered through out the manuscript unless stated otherwise.  %

\section{\label{sec:pr} Distance probability distributions}
\begin{figure}[hbt!] 
	\centering
		{\includegraphics*[width=0.8\linewidth]{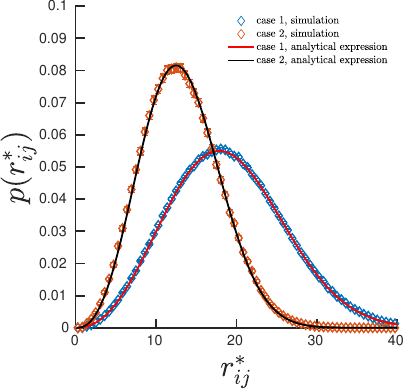}} 		
	\caption{\label{fig:pr} Comparison of distance probability distributions obtained from the simulation of a SAW chain with the analytical expression for case 1 - where both beads are end beads, and case 2 - where one bead is at the chain end and the other bead is an intermediate bead.  }
\end{figure}

The analytical expression given by des Cloizeaux~\cite{des1980short} for the distance probability distribution for a self-avoiding walk (SAW) polymer is
\begin{align}
p( r^*) = C [r^{*}]^{\theta +2} e^{-[K r^*]^{\frac{1}{1-\nu}}}
\end{align}
Here $\nu$ is the Flory exponent, $\theta$ is a geometrical exponent and the coefficients $C$ and $K$ are given by
\begin{align*}
K^2 = \dfrac{\Upgamma([\theta + d+ 2][1-\nu])}{\Upgamma([\theta + d][1-\nu])}\\
%C = 4\pi \dfrac{\Upgamma^{(\theta +d)/2}([\theta + d+ 2][1-\nu])}{\Upgamma^{(\theta +d+2)/2}([\theta + d][1-\nu])}
C = 4\pi \dfrac{[\Upgamma([\theta + d+ 2][1-\nu])]^{\frac{\theta +d}{2}} }{[\Upgamma([\theta + d][1-\nu])]^{\frac{\theta +d +2}{2}} }
\end{align*} 
where $d$ is the dimension. Since our simulations are in 3D, $d = 3$.
The geometrical exponent $\theta$ takes different values in the following three cases 
\begin{enumerate}
	\item Case 1: When both beads are the end beads of the polymer ($\theta=\theta_0$), 
	\item Case 2: When one of the beads is at the end and the other bead is an intermediate bead within the chain ($\theta=\theta_1$),
	\item Case 3: When both the beads are intermediate beads ($\theta=\theta_2$)
\end{enumerate}
As the coefficients $C$ and $K$ depend on $\theta$, they take different values in each of the above cases. 
Following the findings of des Cloizeaux~\cite{des1980short}, Witten and Prentis~\cite{witten1982interpenetration}, Duplantier~\cite{duplantier1989statistical} and Hsu et al. \cite{hsu2004scaling}, one can determine that  $\theta_0= 0.267$, $\theta_1=0.461$ and $\theta_2=0.814~$\cite{santra2021universal}. Simulating a SAW polymer, we compared the probability distribution for all the three cases with the corresponding analytical expressions using the appropriate values of $\theta$. Fig.~\ref{fig:pr} show the validation for case 1 and 2, while the validation for case 3 has been shown in the main text. As can be seen, the simulations are in excellent agreement with the analytical expression. To the best of our knowledge, this is the first comparison of exact numerical results with the analytical expression proposed by des Cloizeaux.

\begin{figure}[hbt!] 
		\centering
	{\includegraphics*[width=0.8\linewidth]{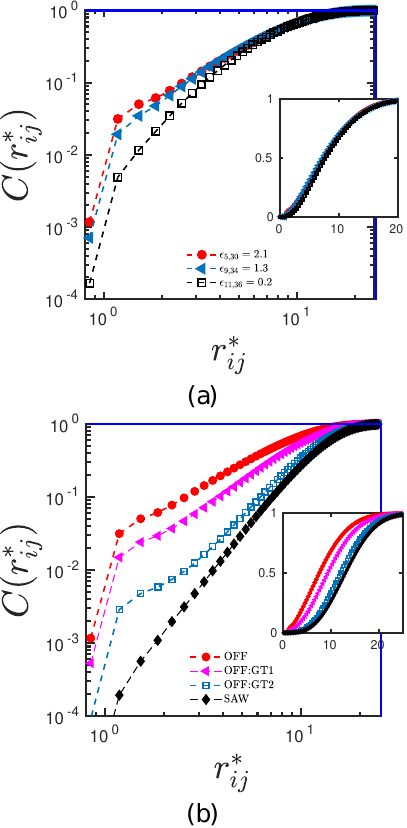}} 
\caption{(a) Cumulative distance distribution $C(r_{ij}^*)$ for various bead-pair  with the same genomic separation $s_{ij} =25$ in the OFF state of $\alpha$-globin gene in log-log scale. The same is indicated in linear scale in the inset. 
	(b) Comparison of $C(r_{ij}^*)$ for the chromatin domain under different ``epigenetic states'' (see text).
	\label{fig:cum_pr}}
\end{figure}

We studied the distance probability distributions of various bead pairs, revealing a distribution with two peaks where one of the peaks is dominated by the entropy of the polymer (genomic separation). The other peak emerges with an increase in the interaction strength. Fig.~\ref{fig:cum_pr}(a) depicts the Cumulative distance distribution $C(r^*)$ for various bead-pairs at same genomic separation ($s_{ij} =25$) experiencing different interaction strengths. Differences in these plots can be easily noticed only at the small $r_{ij}$ while they look similar overall (see inset). The same has been depicted for a specific bead-pair ($5,30$) in different epigenetic states in Fig.~\ref{fig:cum_pr}(b). The difference, in this case, is not only observable for small $r_{ij}$, but for the whole regime of $r_{ij}$ as can be seen in the inset. 

%%%%%%%%%%%%%%%%%%%%%%%%%%%%%%%%%%%%%
\section{\label{sec:epigen} Interaction strength of perturbed states}
\begin{figure*}[hbt!] 
	\centering
	{\includegraphics*[width=0.8\linewidth]{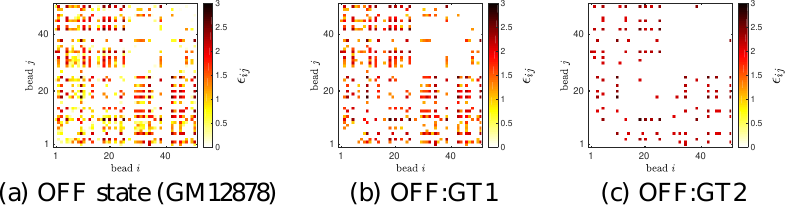}} 
	\caption{\label{fig:epi}(a) The interaction strength obtained form the IBD for GM12878 $\alpha$-globin gene.  (b) A subset of (a) where only the interaction strengths greater than $1k_{\rm B}T$ are considered. (c) A subset of (a) where the interaction strengths greater than $2k_{\rm B}T$ are considered. }
\end{figure*}

The changes introduced into the pair-wise chromatin interactions in the simulations can be considered as equivalent to epigenetic changes. 
In this spirit, we have perturbed the interaction strengths systematically to model different epigenetic-like states. 
The four major states considered in this work are as follows:
\begin{enumerate}
\item OFF - all interactions obtained through the IBD for $\alpha$-globin gene locus in GM12878 (off/repressed state) cell line are considered. These values are obtained in our prior work~\cite{kumari2020computing} and are displayed in Fig.~\ref{fig:epi}(a) with a heatmap representation.  
\item OFF:GT1 - only interactions above $1 k_BT$ are considered (see Fig.~\ref{fig:epi}(b)). All other interactions ($< 1 k_BT$) are taken as $\epsilon_{ij} = 0$ (i.e., there is only steric hindrance between these bead-pairs). This is a subset of the OFF state mentioned in 1. 
\item OFF:GT2 - only strong interactions above  $2 k_BT$ are considered (see Fig.~\ref{fig:epi}(c)). All interactions below $2 k_BT$ are taken as $\epsilon_{ij} = 0$. 
\item SAW - a polymer with only steric hindrance, as a control. In other words, all attractive interactions are switched off. 
\end{enumerate}
Interestingly, Fig.~\ref{fig:epi}(a) and (b) do not look significantly different to the eye, and even in terms of magnitude, only interactions of the order of thermal fluctuation are switched off, yet as demonstrated in the main text, this causes a qualitative change to predictions.

\section{\label{sec:shape} Size and shape analysis }
\begin{figure}[hbt!] 
	\centering
	{\includegraphics*[width=0.8\linewidth]{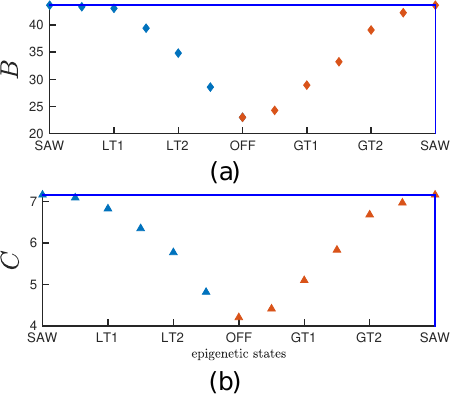}} 
%	\begin{tabular}{c}
%		{\includegraphics*[width=3.in,height=!]{figS6a.pdf}} \\
%		(a)\\
%		{\includegraphics*[width=3.in,height=!]{figS6b.pdf}} \\
%		(b)\\
%		
%	\end{tabular}
	\caption{(a) and (b) represent the asphericity and acylindricity, respectively.
	$x$-axis represents different interaction states with extreme ends representing the control (SAW) polymer and the OFF state (WT) in the middle. LT1 (LT$x$)  indicate that all interactions below 1$k_BT$ ($x k_BT$)are present in the polymer. Similarly GT1 (GT$x$)   indicate that all interactions above 1$k_BT$ ($x k_BT$) are present.  
	\label{fig:shape}}
\end{figure}
Here we discuss various quantities that are used to characterise a chromatin domain.
The radius of gyration of the chain, $R_\text{g} \equiv \sqrt{\langle R_\text{g}^2 \rangle}$, where  $\langle R_\text{g}^2 \rangle$ is defined by 
\begin{equation}
\label{rg}
\langle R_\text{g}^2\rangle=\langle\lambda^2_1\rangle+\langle\lambda^2_2\rangle+\langle\lambda^2_3\rangle
\end{equation}
with, $\lambda^2_1$, $\lambda^2_2$, and $\lambda^2_3$ being the eigenvalues of the gyration tensor $\textbf{G}$ (arranged in ascending order), with
\begin{equation}
\label{eq:gy}
\textbf{G} =  \frac{1}{2N^{2}}\sum_{i=1}^{N}\sum_{j=1}^{N} \textbf{r}_{ij} \textbf{r}_{ij}
\end{equation}
Note that, $\textbf{G}$, $\lambda^2_1$, $\lambda^2_2$, and $\lambda^2_3$ are calculated for each trajectory in the simulation before the ensemble averages are evaluated~\citep{Kuhn1934,Solc1971,Zifferer1999,Haber2000,Steinhauser2005a,Theodorou1985,Bishop1986}.
We examined a large ensemble of configurations and computed shape properties such as
the asphericity ($B$), the acylindricity  ($C$), the degree of prolateness ($S$), and the shape anisotropy ($\kappa^{2}$), as defined below, 
\begin{equation}
B = \langle \lambda^2_{3}\rangle - \frac{1}{2} \left[ \langle \lambda^2_{1}\rangle + \langle \lambda^2_{2}\rangle \right] 
\end{equation}
\begin{equation}
C =  \langle \lambda^2_{2}\rangle - \langle \lambda^2_{1}\rangle
\end{equation}
\begin{equation}
S = \frac{ \left\langle (3\lambda_{1}^2 -  I_{1}) (3\lambda_{2}^2 -  I_{1})(3\lambda_{3}^2 -  I_{1})\right\rangle}{\left\langle \left( I_{1} \right)^{3} \right\rangle}
\end{equation} 
\begin{equation}
\kappa^{2}  = 1 - 3 \frac{\left\langle I_{2} \right\rangle}{\left\langle I_{1}^{2}  \right\rangle}
\end{equation}		
Fig.~\ref{fig:shape}(a) and (b) show the asphericity ($B$) and the acylindricity  ($C$)  for a chromatin domain for various epigenetic states. 
Both $B$ and $C$ increases monotonically as we go from the OFF state to a SAW. 

% \usepackage{graphicx}

\begin{table}[]
\centering
\caption{The values for the shape quantities and $R_g^2$ for the GT and LT case of $\alpha$-globin gene. This data is plotted in Fig.2 of the main manuscript. }
\label{tab:my-table}
\setlength{\tabcolsep}{7pt}
%\resizebox{\textwidth}{!}{%
\begin{tabular}{|c|c|c|c|c|c|c|}
\hline
 & \multicolumn{3}{c|}{LT} & \multicolumn{3}{c|}{GT} \\ \hline
$\epsilon$ & $R_g^2$ & $B/R_g^2$ & $C/R_g^2$ & $R_g^2$ & $B/R_g^2$ & $C/R_g^2$ \\ \hline
0 & 66.3 & 0.658 & 0.107 & 37.6 & 0.612 & 0.183 \\ \hline
0.5 & 66.1 & 0.655 & 0.106 & 39.3 & 0.618 & 0.181 \\ \hline
1 & 66 & 0.652 & 0.103 & 45.8 & 0.631 & 0.173 \\ \hline
1.5 & 60.6 & 0.650 & 0.104 & 51.9 & 0.640 & 0.175 \\ \hline
2 & 54.4 & 0.640 & 0.105 & 59.9 & 0.651 & 0.169 \\ \hline
2.5 & 45.4 & 0.628 & 0.106 & 64.1 & 0.658 & 0.164 \\ \hline
3 & 37.6 & 0.612 & 0.112 & 65.8 & 0.663 & 0.163 \\ \hline
\end{tabular}%
%}
\end{table}

\section{Step-wise convergence of IBD \label{sec:chr7_IBD}}

\begin{figure*}[ht] 
	\begin{center}
			{\includegraphics*[width=1.0\linewidth]{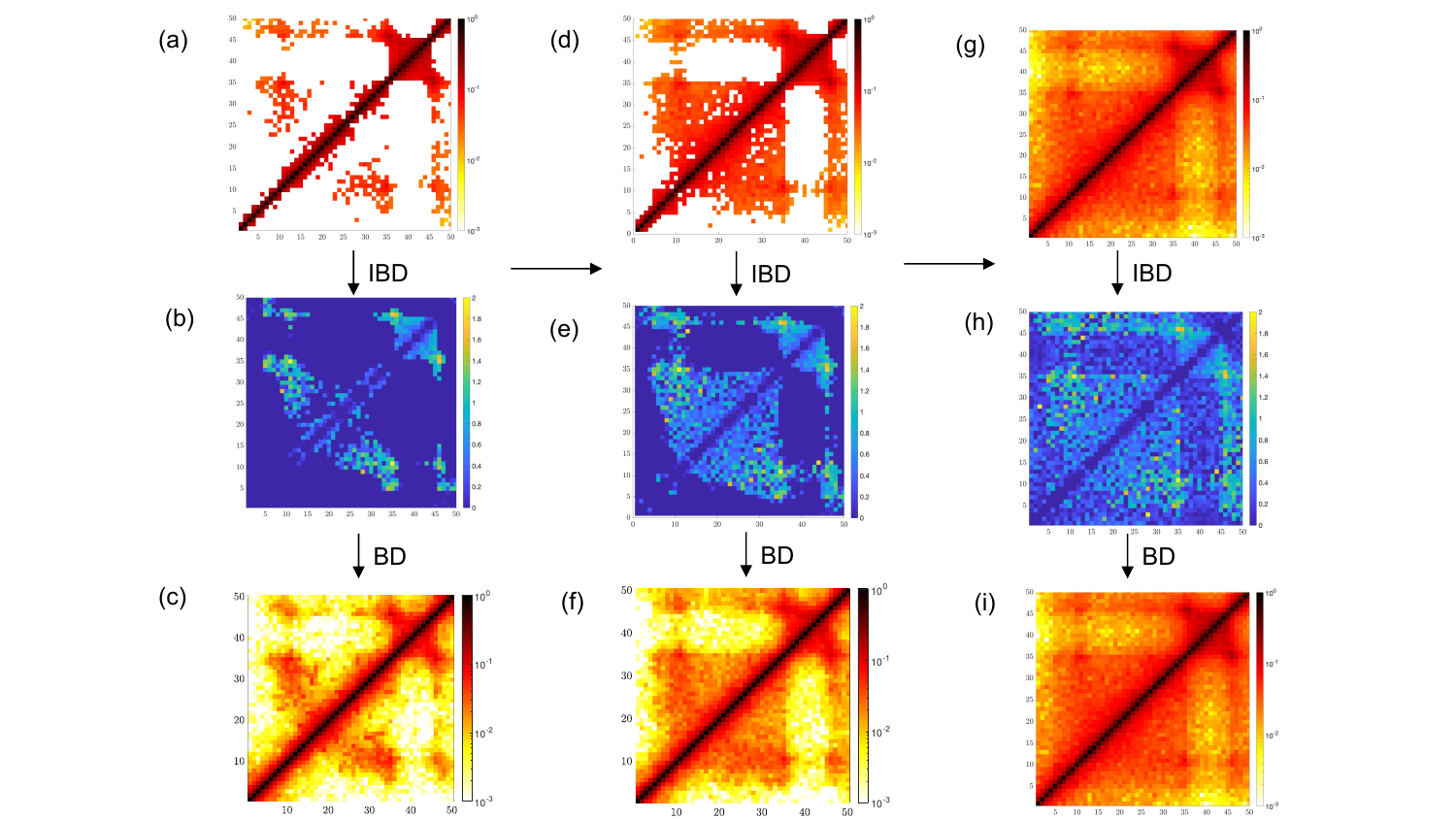}}
		\caption{{\bf Step-wise IBD showing the importance of weak interactions:} Stepwise IBD process for the Chr7 region in the K562 cell line. The IBD optimisation is presented in 3 steps.  (I) First column: input is only the prominent probability values (higher than one standard deviation from the mean. i.e., $p_{ij}>\bar{p}_{|i-j|}+\sigma_{|i-j|}$) show in (a) based on the peak detection algorithm. (b) The corresponding optimised $\epsilon_{ij}$ values, and (c) the recovered contact probabilities. (II) Second column: after the first optimisation step, all peaks above the average probability values (shown in (d)) are fed as input (i.e., $p_{ij}>\bar{p}_{|i-j|}$). The corresponding optimised interaction strengths in (e) are improved with some weak interactions appearing in this 2nd step. However note that the recovered contact probability simulated with prominent interactions in (f) is not comparable to  the full contact probability in (g).  (III) Third column: the complete Hi-C matrix in (g) is fed as input. At the end of this third step  the whole contact probability matrix was recovered in (i), and the corresponding optimal interaction strengths are predicted in (h). \label{fig:k562_peak} }
	\end{center}
\end{figure*}

\begin{figure}[hbt!] 
		\centering
	{\includegraphics*[width=0.8\linewidth]{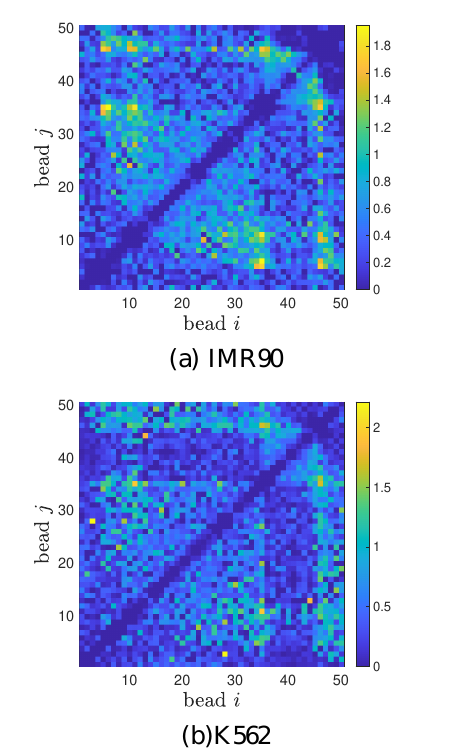}} 
\caption{(a) and (b) show the optimised interaction strength for a domain in IMR90 and K562 cell line, respectively. \label{fig:chr7phi}}
\end{figure}

To classify the contact probabilities into peaks of various strengths, we adopted the following strategy, which we call ``peak detection". Given a segment length, we first calculate the average ($\bar{p}_{|i-j|}$) and standard deviation ($\sigma_{|i-j|}$) of contact probabilities. We then identified the individual contact probability of the bead pair as a prominent peak if the contact probability is greater than the summation of average and standard deviation of contact probability at that segment length ($p_{ij}>\bar{p}_{|i-j|}+\sigma_{|i-j|}$). Similarly, we identified the intermediate peaks if the contact probability is greater than the average contact probability at that segment length ($p_{ij}>\bar{p}_{|i-j|}$). For the stepwise optimization of IBD, we first optimized only for the prominent peaks. Once the prominent peaks are recovered, we optimized for the intermediate peaks. In the end, we optimize for the full contact probability matrix. For instance, Fig.~\ref{fig:k562_peak} shows the stepwise IBD optimization of the domain in the K562 cell line. Similar to the case for IMR90 discussed in the main text, here, the first row is for the prominent peaks showing the reference contact probability on top, optimized interaction strength at the middle and recovered contact probability at the bottom. Similarly, the second and third column is for the intermediate and full contact probability matrix. Fig.~\ref{fig:chr7phi}(a) and (b) show the optimised interaction strength for the IMR90 and K562 cell line. The step-wise convergence process makes the convergence of the whole domain easier; i.e., it is easier to obtain  the correct interaction strength values that would reconstruct the experimentally observed contact maps. Given that the landscape is complex with possible metastable states/barriers, it is not guaranteed that the system will converge to a HiC-like state within a reasonable simulation time. A naive effort to converge accounting for contacts of all bead-pairs often get stuck in states with large error values and fails to converge.
%fetch the optimal interaction strengths that would reproduce the observed contact map. }
 %\tr{The step-wise convergence process reduces the risk of divergence in the system and lead to the correct interaction strength values which would reconstruct the experimental contact maps. A simple convergence for all bead-pairs at once might lead to complexity and often result in divergence of the system and fails to fetch the interaction strength which would reproduce the experimental contact map.}

\begin{figure}[hbt!] 
		\centering
	{\includegraphics*[width=0.8\linewidth]{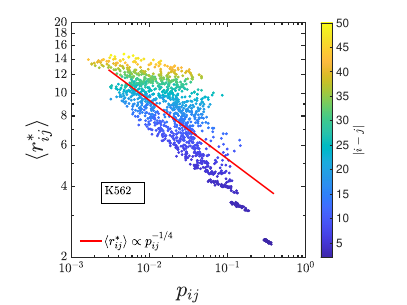}} 
\caption{Our prediction of the mean 3D distance between every pair of beads ($r_{ij}$) as a function of their corresponding contact probability ($p_{ij}$) for cell line K562. The color indicates genomic distance $|i-j|$ between the pair of beads (in units of bead size, see sidebar). Curve with a power-law relation $r_{ij} \propto p_{ij}^{(-1/4)}$ is plotted (red line)  as a guide to the eye. 
\label{fig:SI_3dpij}}
\end{figure}

After the convergence of the IBD, we simulated the system with optimal interaction strengths and investigated the relation between the contact probability and the spatial distance between all segment pairs of chromatin. Fig.~\ref{fig:SI_3dpij} shows our prediction of the mean 3D distance between every pair of beads ($r_{ij}$) as a function of their corresponding contact probability ($p_{ij}$) for cell line K562. One can observe that there is a broad distribution of 3D distances around the mean value. 

%\section{Contact probability and 3D distance}

\section{\label{sec:temporal} Quantifying the temporal nature of chromatin domains}
To study the dynamics of the chromatin domain, we quantified several temporal quantities such as relaxation time, loop formation time and contact time. Temporal variation in properties of chromatin 
can be found by observing the spatial distance between two segments as a function of time.

\begin{figure}[hbt!] 
	\centering
	{\includegraphics*[width=0.8\linewidth]{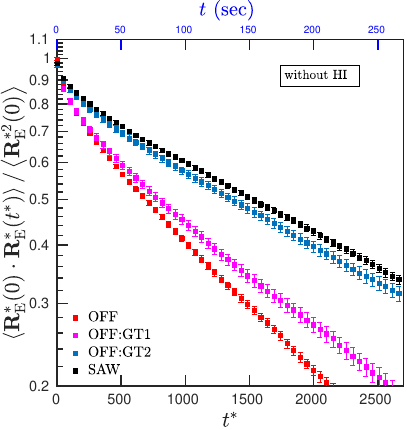}} 
	\caption{Exponential decay of end-to-end auto-correlation function with time. Here the HI is not included in the simulation. 
	\label{fig:autocor}}
\end{figure}

As a first step towards quantify dynamics, we extracted the longest relaxation time from the end-to-end vector autocorrelation function $\left<\mathbf{R}_{\mathrm{E}}^*(0)\cdot\mathbf{R}_{\mathrm{E}}^*(t^*)\right>/\left<\mathbf{R}^{*2}_{\mathrm{E}}(0)\right>$ where $\mathbf{R}_{\rm E}^* = |r_1^*-r_{50}^*|$. This computes the time-dependent correlation of the end-to-end vector. What it essentially measures is how long does it take for the end-to-end beads to fluctuate around its equilibrium value. Fig.~\ref{fig:autocor} displays the autocorrelation function decay with time not accounting for HI. The main paper shows the same with HI and the extracted relaxation time for both (with and without HI). 
In the OFF state, intra-chromatin interactions are high compared to the SAW; as can be seen from the $p(r^*)$ (Fig.1(c)), in the OFF state, there is a sharp peak in $p(r^*)$ (equivalently there is a steeper minima in free energy, inset of Fig.6(c)) suggesting that the end-to-end beads cannot fluctuate too much. Our analysis of the effective stiffness (Fig.6(c)) further shows that, in the OFF state, the end-to-end points can be imagined as two beads connected by a stiff effective spring (arising from interactions). The spring in the OFF state is stiffer compared to the weaker entropic spring in the SAW. Two points connected by a stiffer spring will have a lower fluctuation time scale (inversely proportional to the effective stiffness). Hence we get a lower timescale in our calculation.
Another way of thinking is that the relaxation time will be proportional to $R_g^2$. For the OFF state, $R_g$ is small leading to a smaller relaxation time.

\begin{figure}[hbt!] 
	\centering
	{\includegraphics*[width=0.8\linewidth]{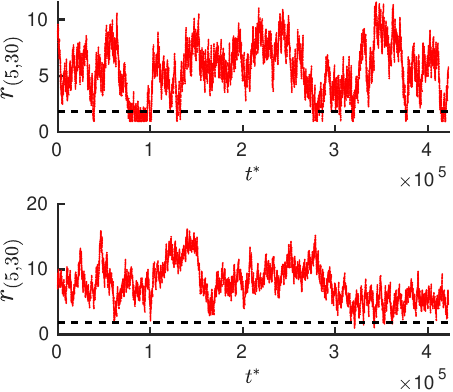}} 
	\caption{Distance data from our simulation for a particular pair of beads for two randomly chosen realisation.
	\label{fig:traj}}
\end{figure}

\begin{figure}[hbt!] 
	\centering
	{\includegraphics*[width=0.8\linewidth]{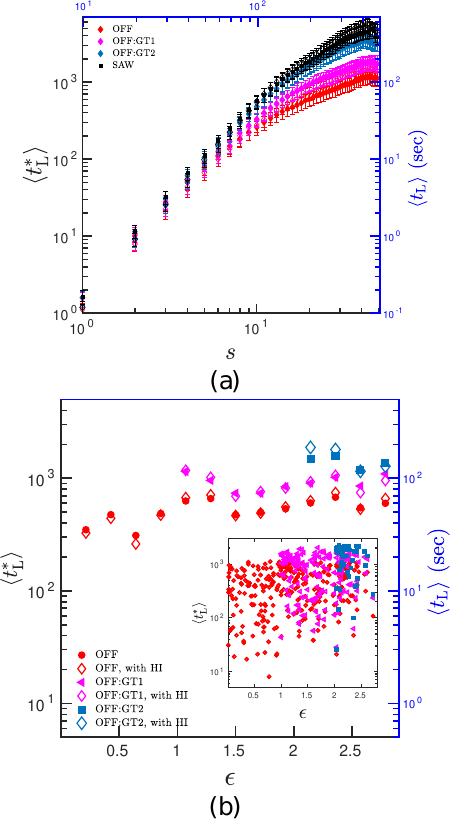}} 
	\caption{(a) $\langle t_{\rm L}^* \rangle$ has a power law scaling with genomic length ($\langle t_{\rm L}^* \rangle \sim s^{\mu} $). Simulations with HI (empty symbols) and without HI (filled symbols) falls on top of each other and are indistinguishable.
			(b)  $\langle t_{\rm L}^* \rangle$ binned and averaged over all bead pairs having same $\epsilon$ showing minimal influence of $\epsilon$. 
			Inset: $\langle t_{\rm L}^* \rangle$ as a function of $\epsilon$ with each point representing a bead pair. 
	\label{fig:tl}}
\end{figure}

We then looked at the loop formation time ($t_{\rm L}^*$) defined as the time taken for a pair of beads to meet ($r_{ij}^* < r_{\rm C}^*$) for the first time, starting from a random equilibrium configuration. 
Fig.~\ref{fig:traj} showcases the spatial distance between a specific bead pair (bead $5$ and bead $30$) with time. It can be easily observed that within a single trajectory, bead-pairs come in contact 
several times.
We also looked at the parameters affecting average loop formation time ($\langle t_{\rm L}^* \rangle$). Fig.~\ref{fig:tl}(a) and (b) depict $\langle t_{\rm L}^* \rangle$ with genomic separation ($s$) and interaction strength ($\epsilon$), respectively. All the filled symbols in Fig.~\ref{fig:tl} represent results from simulation, including HI, while the empty symbols represent no-HI case. 
The study shows that while the HI is crucial in determining the relaxation time of the whole polymer, it does not affect temporal quantities such as $\langle t_{\rm L}^* \rangle$. 

\begin{figure}[hbt!] 
	\centering
	{\includegraphics*[width=0.8\linewidth]{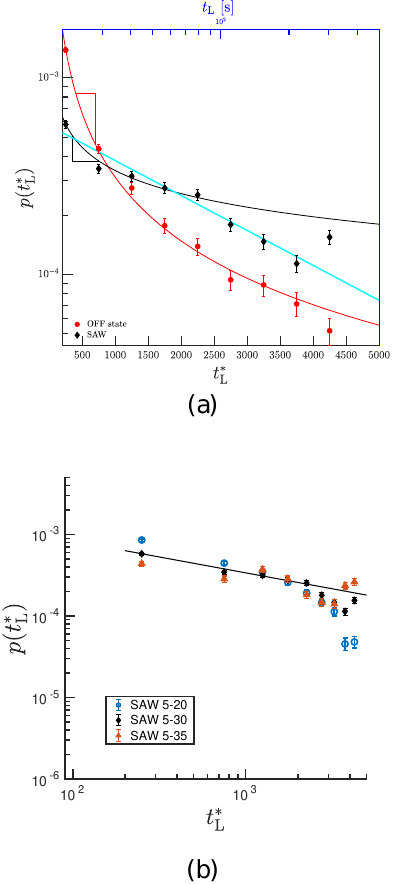}} 
		\caption{{(a)Fitting power-law and exponential function to examine the nature of distribution function for loop formation time ($t_{\rm L}^*$) for a specific bead-pair (bead 5 and bead 30) in SAW and OFF state. A power-law function fits to the early part of SAW and the exponential function fits better to the late part. This is expected because, due to the finite size of the system, powerlaw cannot extend indefinitely; For late times, it should decay quickly as the typical time for two beads, $L$ length away, to loop cannot be too large. This is a typical finite size effect.
	(b)Probability distribution of loop formation time showing the effect of finite chain size in SAW. Smaller segment length such as 5-20 deviates from the power law (black solid line) much sooner compared to the larger segment lengths like 5-35.}
	\label{fig:ptl}}
\end{figure}

We also examined the distribution of $t_{\rm L}^*$ which shows a power law nature with exponential tails (see Fig.~\ref{fig:ptl}(a) and (b)). The exponential tail is a typical finite size effect. Since chromatin is of finite size, this is expected. 


~\\
\noindent
{\bf Relation between contact time and interaction energy}\\
~\\
Fig.8(b) in the main text shows a relation between contact time and interaction energy as $t_c \propto \exp{(\epsilon/2)}$. However, one would naively expect that the inverse of the time -- the rate of breaking the contact -- would be proportional to $e^{-\epsilon}$ from the perspective of a simple Kramers' problem. However, note that this expectation is true only if we assume a Kramers' problem for a single particle in a simple potential well with a barrier of height $\epsilon$. Here, even though the $\epsilon$ is the interaction energy between two beads, there is a whole polymer influencing both binding and dissociation. Even if we assume a simple potential well, the only thermodynamic constraint is that the ratio of the binding and dissociation rates should be $e^{\Delta G}$, where $\Delta G$ has contributions from both energy and entropy. Here $\epsilon$ is just the energy part. The ratio can also have an entropy contribution. Hence, it may deviate from $e^{-\epsilon}$.

From another point of view, the true energy landscape need not be a simple one. We do not know the precise (multi-dimensional) free energy landscape; hence, individually, the rates (times) could have a factor of 2 (or any factor for that matter which may not be predictable apriori).

\newpage
